\documentclass[twocolumn]{aastex63}
\usepackage{amsmath} \usepackage[normalem]{ulem} \usepackage{xcolor}
\usepackage{url} \usepackage{cancel}

\catcode`_=\active \newcommand_[1]{\ensuremath{\sb{\mathrm{#1}}}}
\newcommand{\figgris}
{\begin{figure*}
\includegraphics[width=0.33\linewidth]{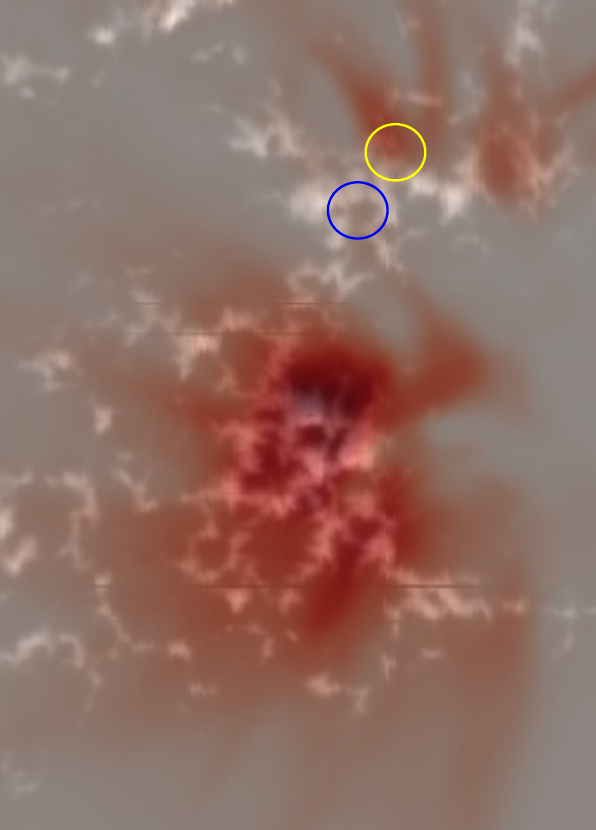}	
\includegraphics[width=0.33\linewidth]{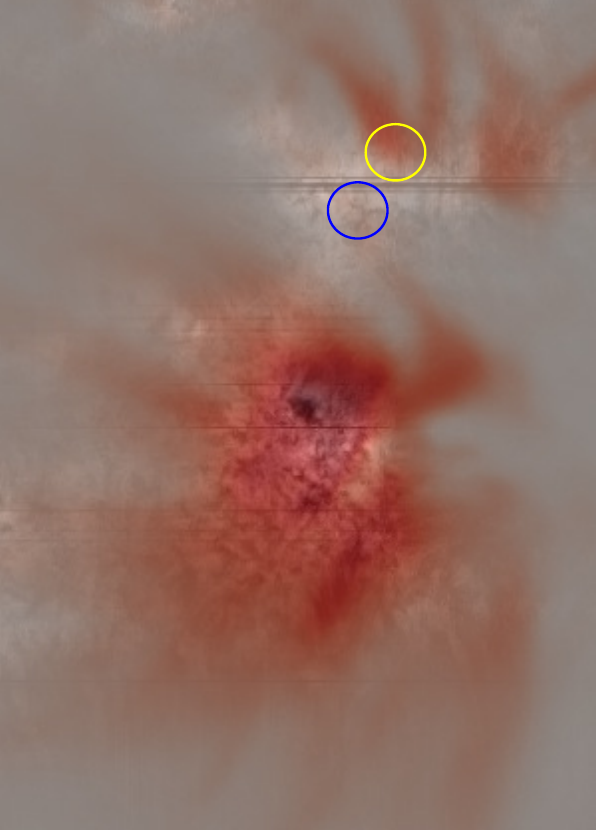}	
\includegraphics[width=0.33\linewidth]{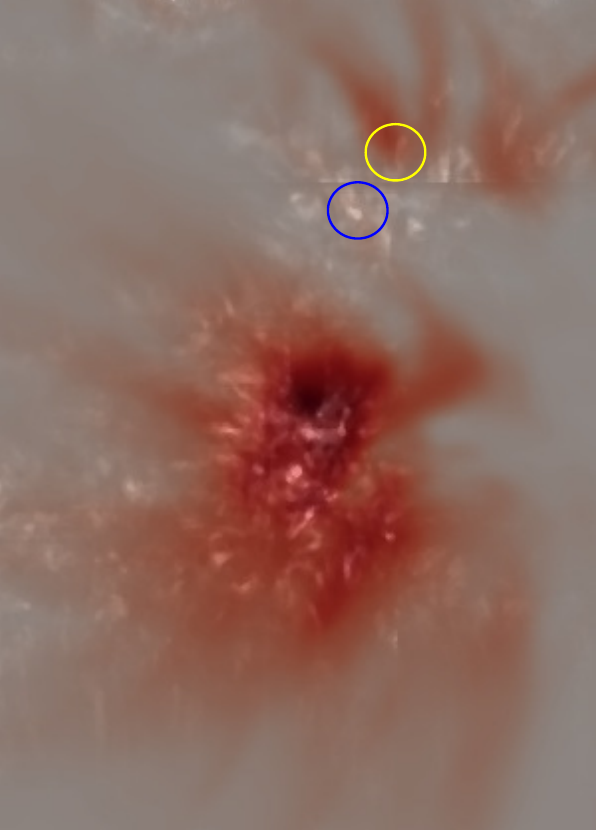}	
\caption{Line of sight 
photospheric magnetic fields (left column), chromospheric magnetic fields (middle), 
and scalar proxy (expression 7) of Poynting flux (right)  are rendered in 
grayscale for data from the GRIS instrument at the GREGOR telescope, on 28 September 2020 
(see J24b for details). The region is centered near 
$X=-13$, $Y=325$ in heliographic
coordinates.  
The photospheric line of \ion{Si}{1} $\lambda 1082.5$ nm, and chromospheric blended lines of 
\ion{He}{1} at 1083.0 nm 
were used to derive the
quantities shown.
Superposed in red are<
negative images of
the footpoints of coronal loops as seen 
in the AIA 17.1 nm channel, which have very similar morphology to
other AIA channels (J24b).
The 
field of
view is  
$40\farcs5\times 60\farcs5$, set by the area scanned by the GRIS slit.   The yellow circled
region is the footpoint
of a hot plasma loop base,
the otherwise similar underlying region, circled in blue, is not.
\label{fig:gris}
}
\end{figure*}
}

\newcommand{\figbrwjv}
{\begin{figure}
\includegraphics[width=\linewidth]{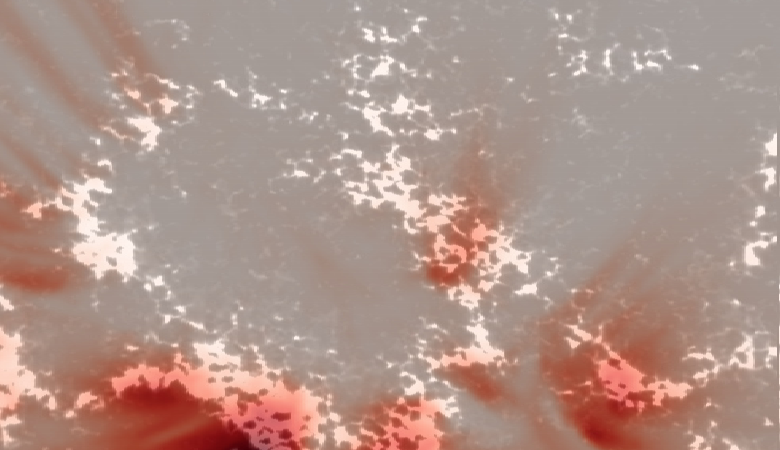}	
\includegraphics[width=1.0\linewidth]{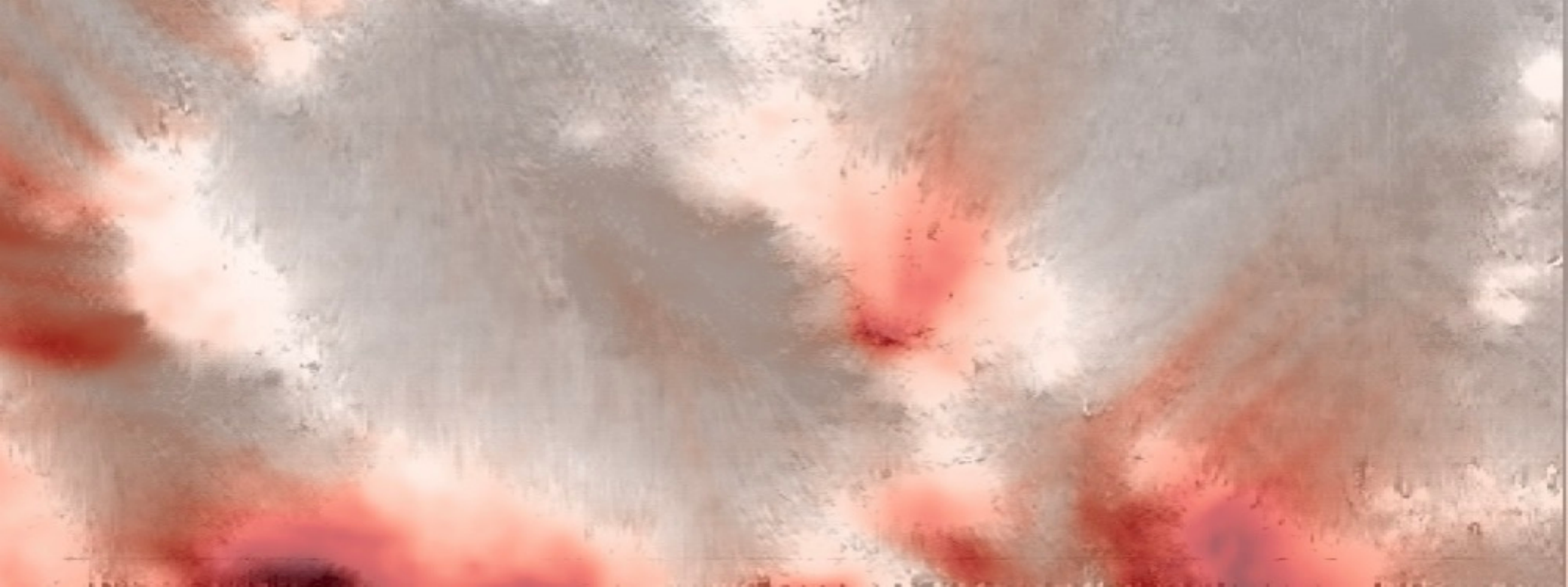}	
\includegraphics[width=1.0\linewidth]{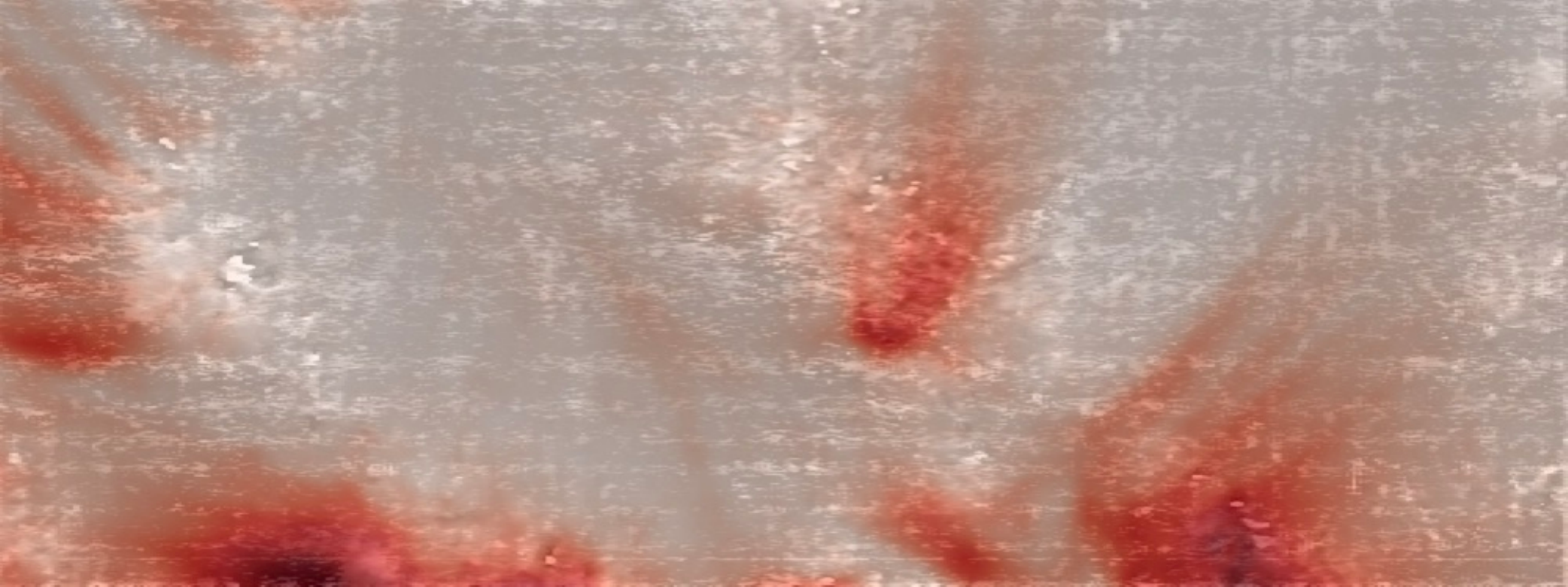}	
\caption{Line of sight 
photospheric magnetic fields (top row), chromospheric magnetic fields (middle row), 
and scalar proxy of Poynting flux (bottom
row, from expression 7) are rendered in 
grayscale for DKIST dataset 
BRWJV+AVORO  (J24a),
two scans centered at 3-Jun-2022 17:52:23 at $X=-411$, $Y=377$.
The photospheric lines of \ion{Fe}{1} near $6302$ nm, and chromospheric  line of 
\ion{He}{1} at 854.2 nm 
were used to derive the
quantities shown.
Superposed in red are
negative images of
the footpoints of coronal loops as seen 
in the AIA 17.1 nm channel.  Owing to the 
different magnifications 
of the two arms of
the ViSP spectrograph
used, the photospheric
field has a field of
view of $100\arcsec\times77\arcsec$, the other images
$100\arcsec\times50\arcsec$.
\label{fig:brwjv}
}
\end{figure}
}

\newcommand{\figaodmm}
{\begin{figure}
\includegraphics[width=1.0\linewidth]{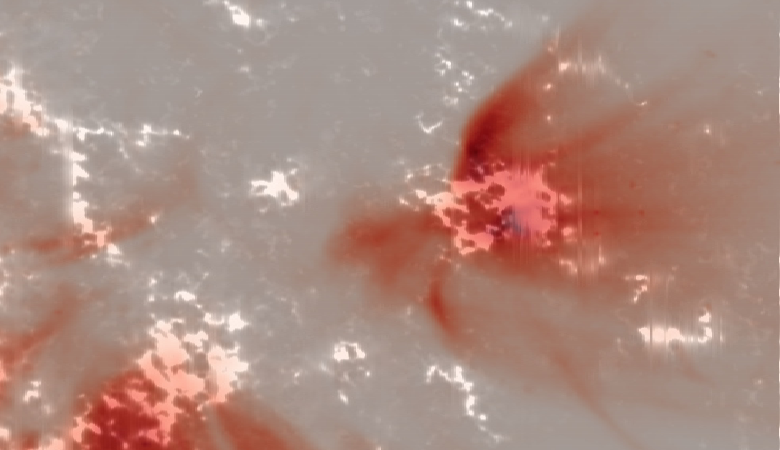}	
\includegraphics[width=1.0\linewidth]{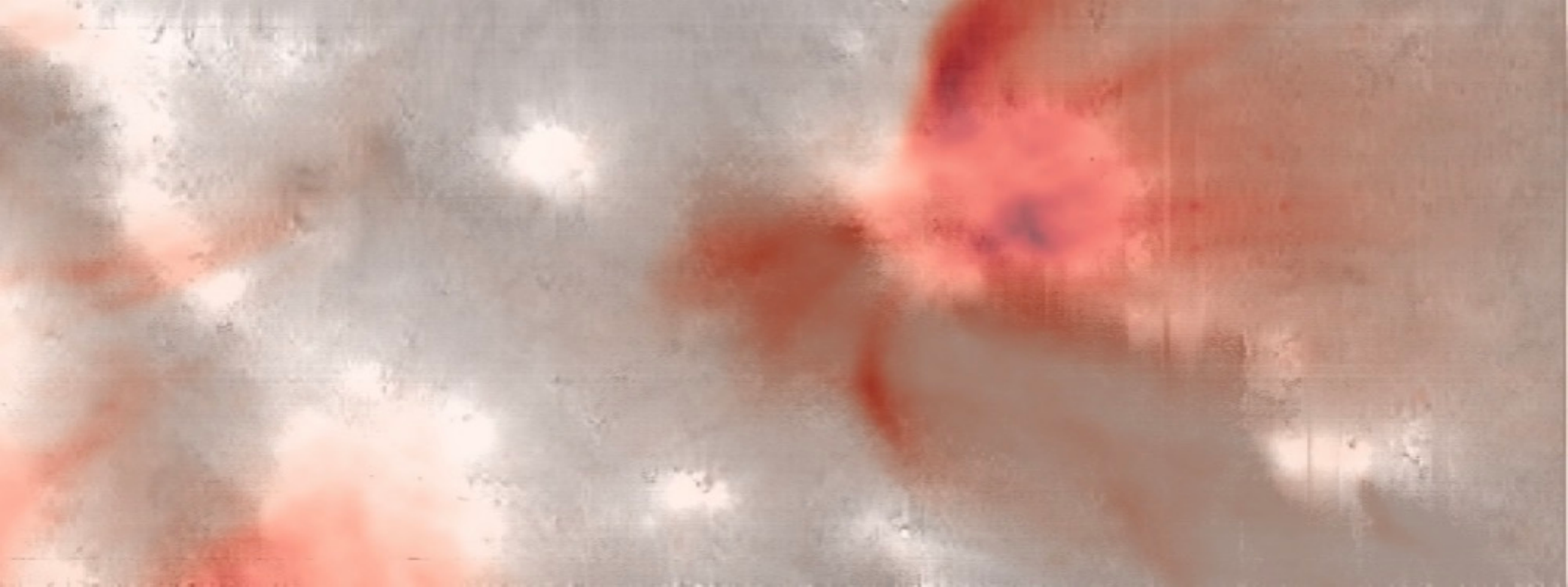}	
\includegraphics[width=1.0\linewidth]{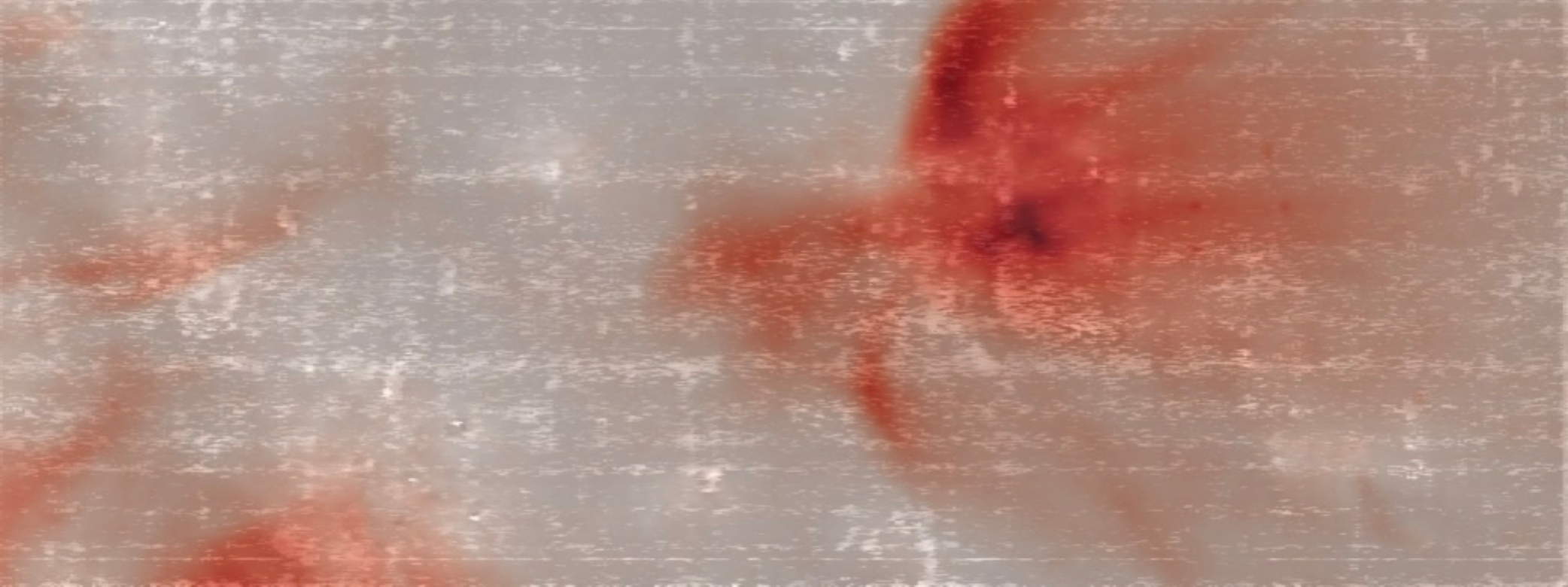}	
\caption{Magnetic and 
coronal properties are
shown as in Figure~\ref{fig:brwjv},
except for DKIST dataset AODMM+APJND. The mid time of the scan 
was 2-Jun-2022 20:02:23, it was centered near $X=-495$, $Y=495$. 
}\label{fig:aodmm} 
\end{figure}
}

\newcommand{\figawowp}
{\begin{figure}
\label{fig:awowp} 
\includegraphics[width=1.0\linewidth]{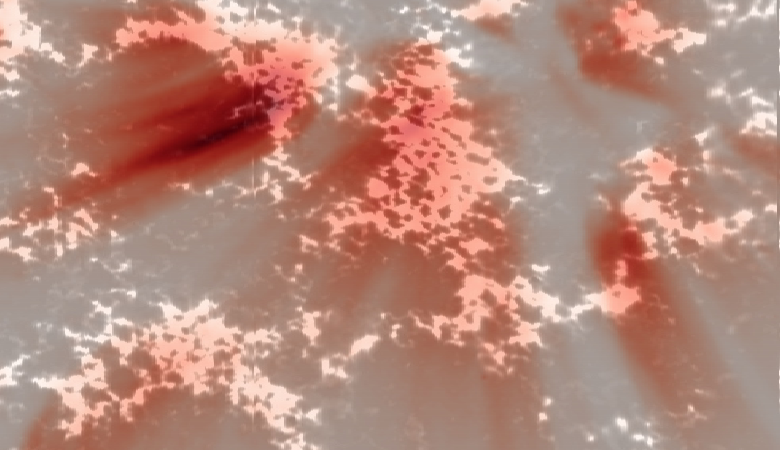}	
\includegraphics[width=1.0\linewidth]{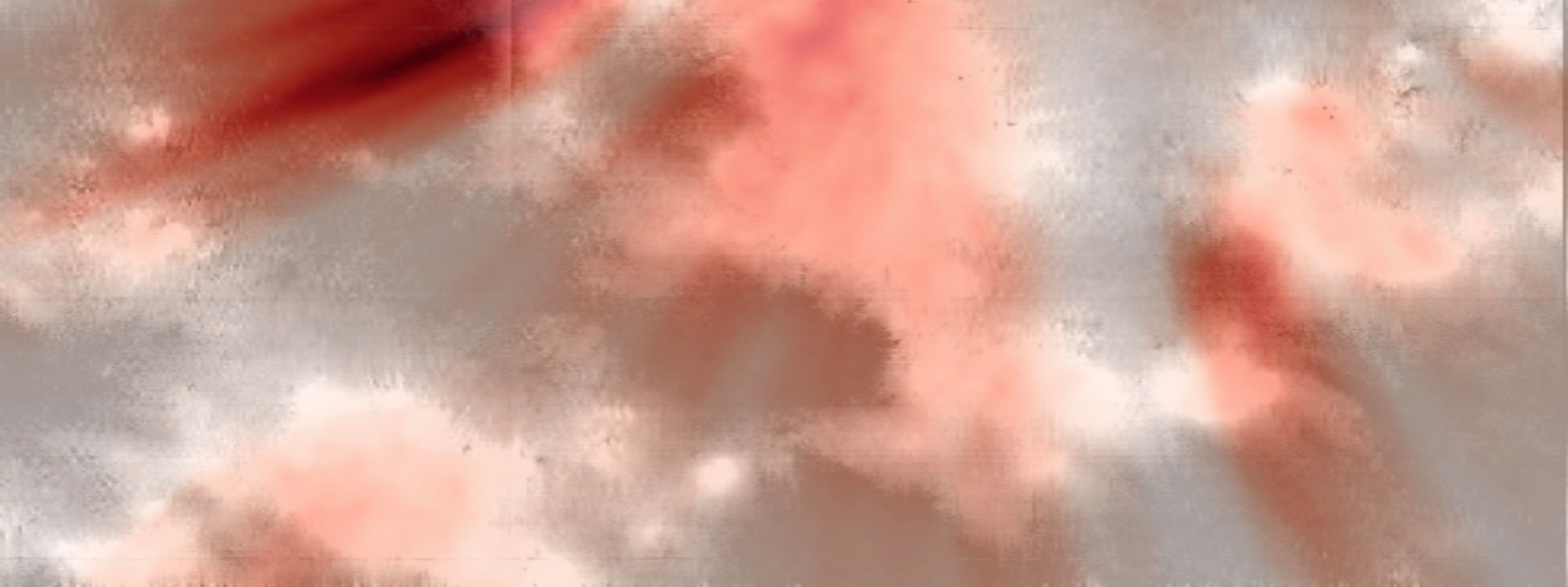}	
\includegraphics[width=1.0\linewidth]{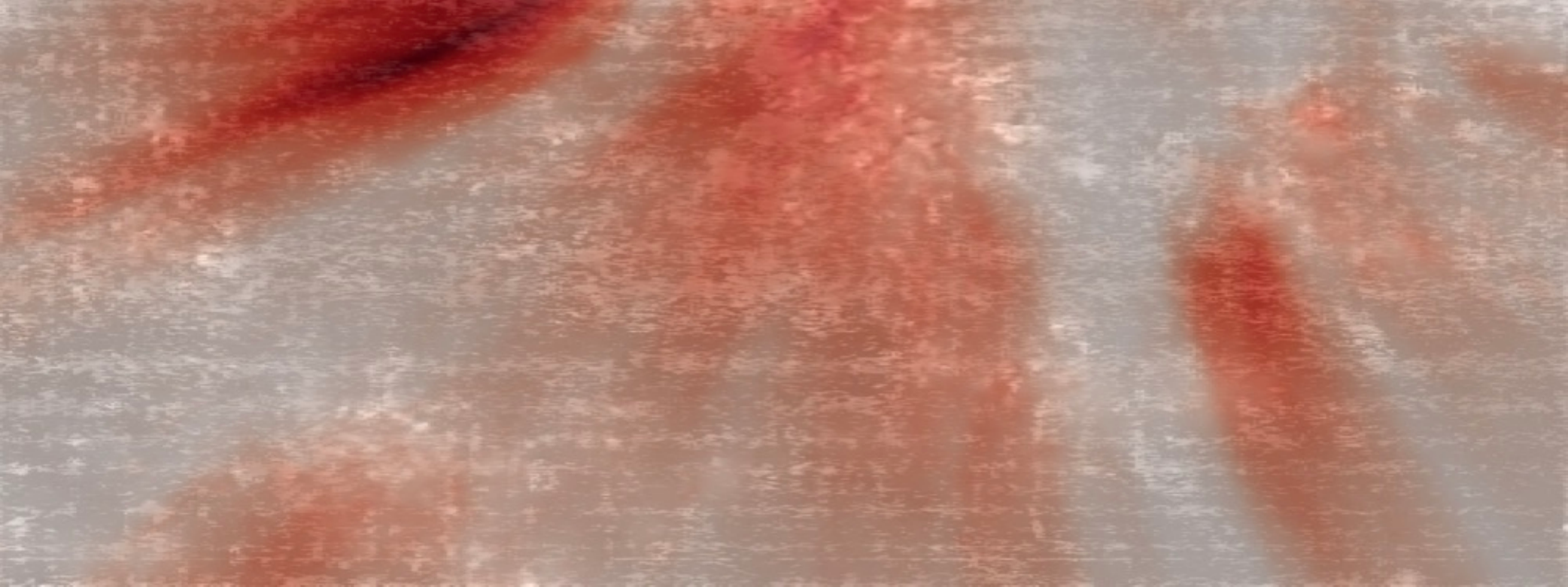}	
\caption{Magnetic and 
coronal properties are
shown as in Figure~\ref{fig:brwjv},
except for scans AWOWP+AXVLY, 
with a mid time of 23 June 2023
17:23:21, centered near $X=-416$, $Y=-436$.
}
\end{figure}
}

\newcommand{\figbpjdd}
{\begin{figure}
\includegraphics[width=1.0\linewidth]{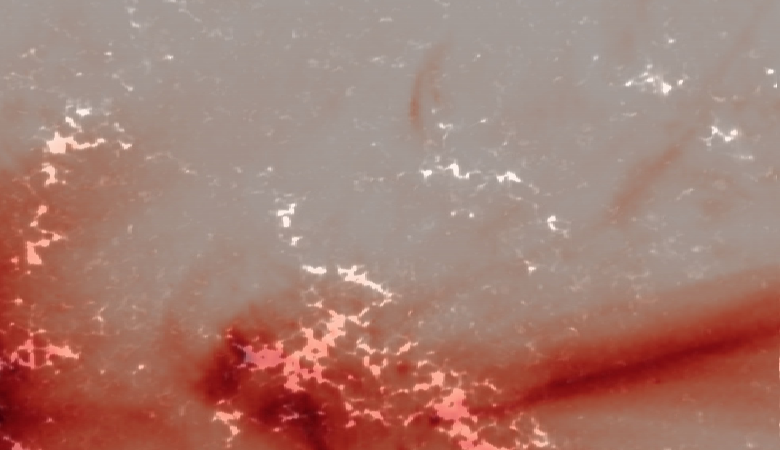}	
\includegraphics[width=1.0\linewidth]{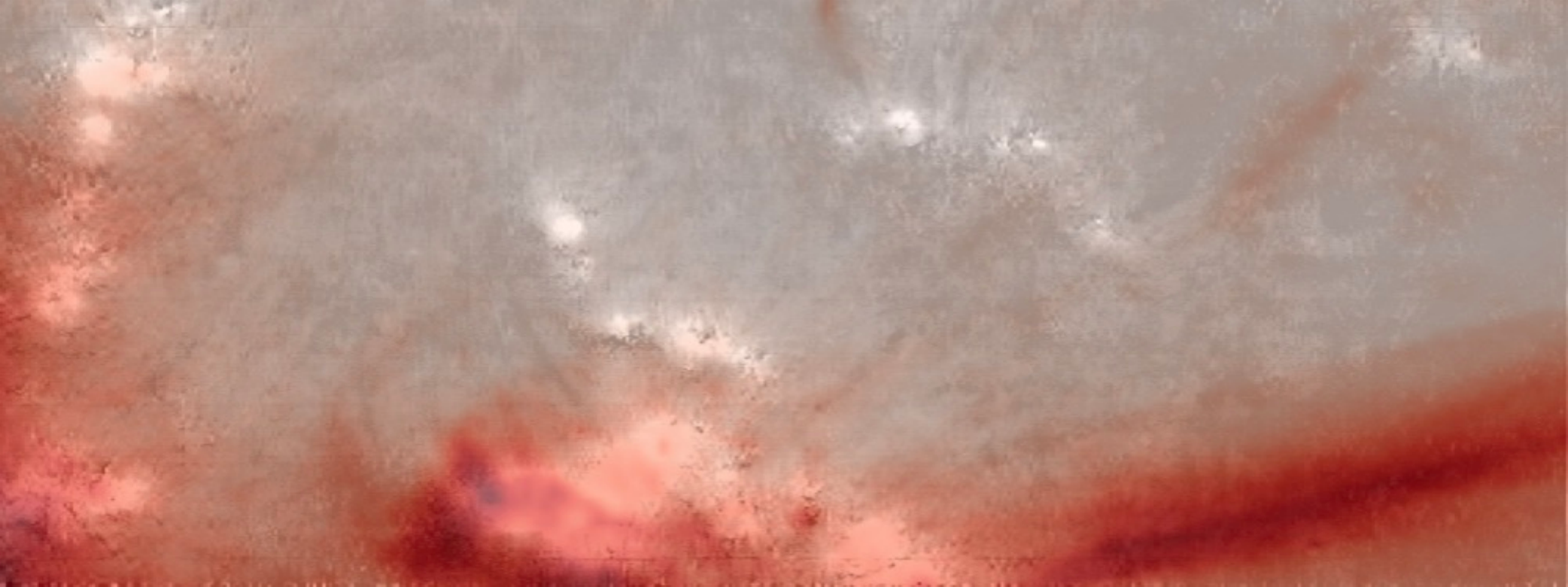}	
\includegraphics[width=1.0\linewidth]{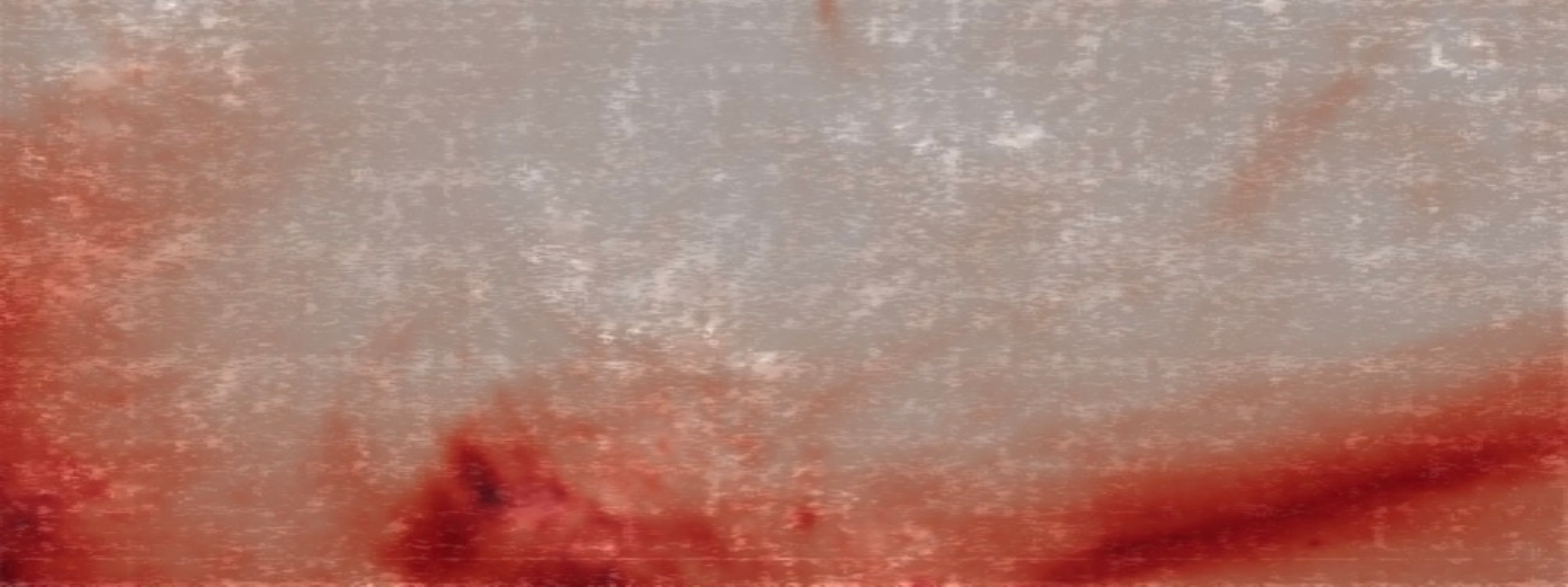}	
\caption{Magnetic and 
coronal properties are
shown as in Figure~\ref{fig:brwjv},
except for DKIST dataset BPJDD+BQKZZ, obtained at midtime 18:21:07 on Jun 23 2023.  $X=-312$, $Y=-377$.
} \label{fig:bpjdd} 
\end{figure}
}

\newcommand{\figbnkvm}
{\begin{figure}
\includegraphics[width=1.0\linewidth]{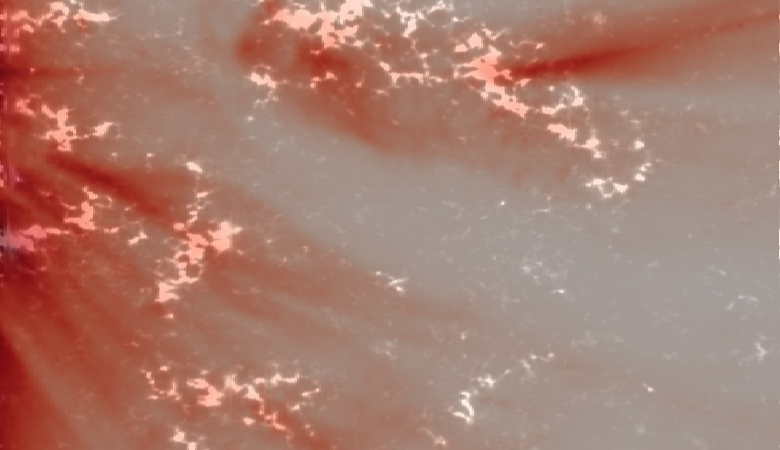}	
\includegraphics[width=1.0\linewidth]{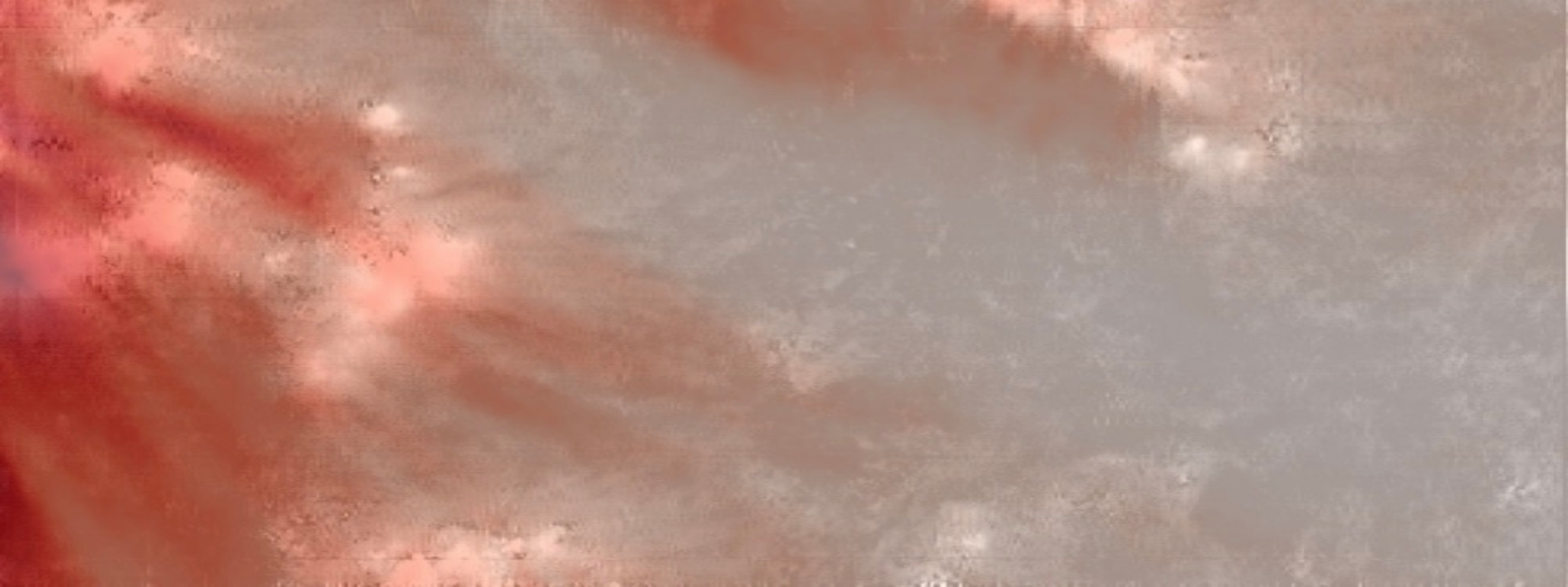}	
\includegraphics[width=1.0\linewidth]{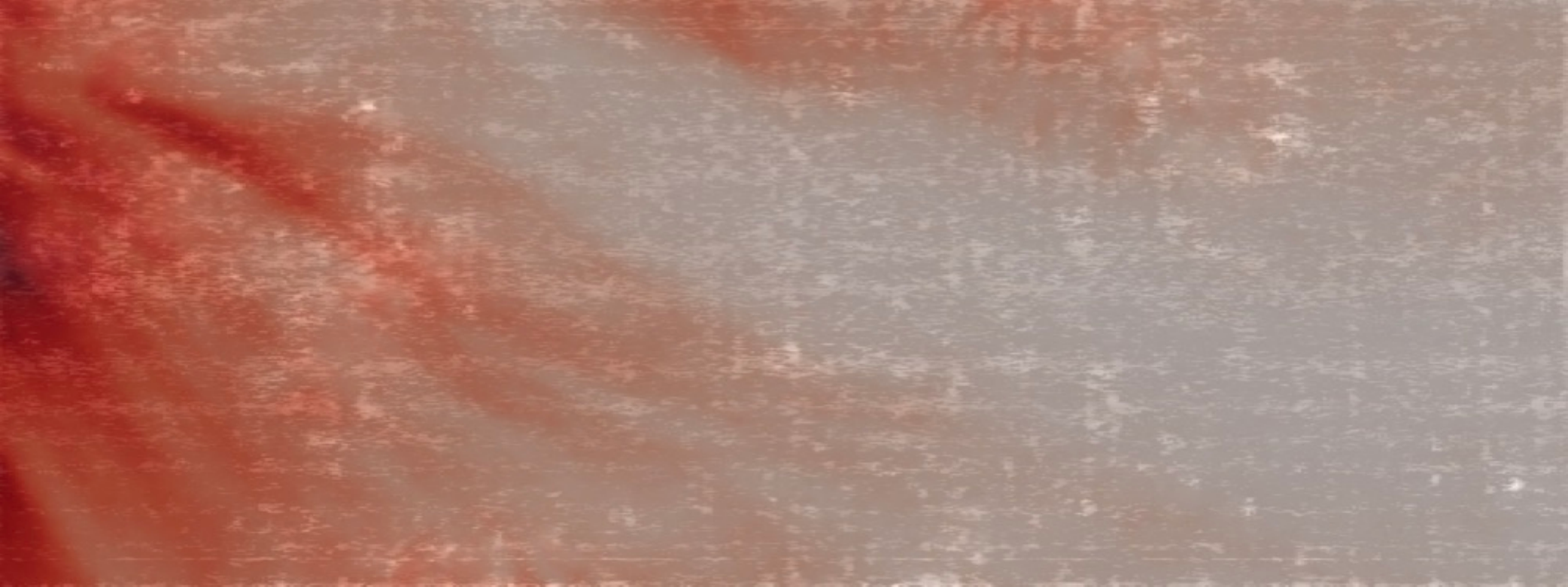}	
\caption{Magnetic and 
coronal properties are
shown as in Figure~\ref{fig:brwjv},
except for dataset DKIST BNKVM+BODXM, obtained at midtime 18:49:52 
 on Jun 23 2023, centered at  $X=412$ $Y=439$.
} \label{fig:bnkvm} 
\end{figure}
}

\newcommand{\hao}{
High Altitude Observatory,
National Center for Atmospheric Research,
Boulder CO 80307-3000,
 USA}

\shorttitle{} \shortauthors{Judge \& Kuin}

\begin{document}

\title{On the intermittency of hot plasma loops in the solar corona}


\correspondingauthor{Philip G. Judge} \author{Philip G. Judge}
\affiliation{Visitor, Astronomical Institute of the University of
  Bern, Sidlerstrasse 5, 3012 Bern, and\\ \hao}

\author{N. Paul M. Kuin} \affiliation{ Mullard Space Science
  Laboratory, University College London, Holmbury St. Mary, Dorking
  RH5 6NT, UK}
\date{Accepted . Received ; in original form }


%
%

\begin{abstract}
A recent analysis has suggested that the heating of plasma loops in
the solar corona depends not just on the Poynting flux but also on
processes yet to be identified.  This discovery reflects and refines
earlier questions such as, why and how are entire hydromagnetic
structures only intermittently loaded with bright coronal plasma
\citep{Litwin+Rosner1993}?  The present work scrutinizes more
chromospheric and coronal data, with the aim of finding reproducible
observational constraints on coronal heating mechanisms.  Six
independent scans of chromospheric active region magnetic fields are
investigated and correlated to overlying hot plasma loops. For the first time, the
  footpoints of over 30 bright plasma loops are thus related to scalar
  proxies for the Poynting fluxes measured from the upper
  chromosphere.  Although imperfect, the proxies all indicate a
  general lack of correlation between footpoint Poynting flux and loop
  brightness.  Our findings consolidate the claim that unobserved
physical processes are at work which govern the heating of long-lived
coronal loops.
\end{abstract}


\section{Introduction}
\label{sec:statement}

Sometimes in a mature research field, there is a need to step back and
re-assess some fundamental facts.  The present paper aims to establish
a clearer observational picture of the physical connection between the
magnetic field near the coronal base, particularly the Poynting flux,
and bright coronal plasma.  {For decades correlations have been
  found between photospheric magnetic fluxes and X-ray brightness of
  the corona \citep[examples include ][]{Krieger+others1971,
    Poletto+others1975, Golub+others1980,
    Schrijver+harvey1989,Schrijver+Zwaan2000,
    Pevtsov+others2003,Barczynski+others2018,
    Toriumi+Airapetian2022}.}  It has long been established that the
solar X-ray corona above active regions consists of multiple loop-like
structures, heated somehow through the magnetic fields confining them
\citep[e.g.][]{Vanspeybrock+Krieger+Vaiana1970,Rosner+Tucker+Vaiana1978}.
These are now believed to consist of ``tubes'' of hot plasma, with
amorphous cross sections, aligned within magnetic flux surfaces and
extending upwards from strongly magnetized regions in the lower
atmosphere \citep{Malanushenko+others2022}.  The simple fact that we
clearly see such structures begs the question of what organizes the
plasma into loops on observable scales
(\citealp{Rosner1990,Litwin+Rosner1993}. The latter we refer to as
``LW93'').  Why are certain tubes, {parts of large hydromagnetic
  structures, so} full of hot plasma and others not?

\figgris \figbrwjv

{This problem presents potential challenges for typical thermal
  models which, for a given upward flux of electromagnetic energy
  dissipated in the corona, produce unique and deterministic
  equilibria \citep[see the intriguing history of 1D and 0D modeling
    in articles by][]{Kuin+Martens1982,McClymont+Craig1985a,
    McClymont+Craig1985b,Craig+Schulkes1985,
    Vandenoord+Barstow1988,Craig1990,Martens2010,
    Cargill+Bradshaw+Klimchuk2012b}.
  Under the current picture of coronal heating, dissipative structures
  exist only on small scales below $\sim 200$ km, perhaps approaching
  kinetic scales of tens of meters \citep{Judge+Ionson2024}.  Some
  have proposed, based in part on this picture, that} unresolved
threads or ribbons of plasma within loops dominate the emission from
within each loop \citep{Aschwanden+others2000,
  Patsourakos+Klimchuk2005,
  Klimchuk2006,Aschwanden+others2007,EBTEL,Lopez+Klimchuk2022}.
{But then one faces the question of why the heated plasma
  simultaneously exists both on fine scales, yet is also organized on
  the larger observable scales, which currently exceed $\sim 200$ km
  at EUV wavelengths \citep{Rachmeler+2022}.  Is the upward Poynting
  flux organized on such scales? Recent findings of
  \citet{Judge+others2024a}, henceforth ``J24a'', based upon data of
  one active region, suggest not.  Instead these authors speculated
  that the corona itself might regulate its own rate of heating, an
  idea with origins in some theoretical work
  \citep{Uzdensky2007,Einaudi+others2021}. }

J24a analyzed chromospheric magnetic fields using the ViSP instrument
\citep{Dewijn+others2022} to measure states of polarization of the
854.2 nm line of \ion{Ca}{2} at the Daniel K. Inouye Solar Telescope
\citep[``DKIST'',][]{Rimmele+others2020}.  These data were compared
with space-based measurements of overlying coronal plasmas.
Quantities related to the upward Poynting flux at the base of coronal
plasma loops {were found to be essentially uncorrelated} with the
brightness of footpoints of heated plasmas observed with the EUV
channels of the AIA instrument \citep{Lemen+others2012} on SDO
\citep{Pesnell+others2012}.  {Such a result, if confirmed, would
  represent a new fundamental fact with which to test coronal heating
  mechanisms.}

{The main purpose of the present paper is to find further
  observational evidence for or against this result.  Our approach is
  to use chromospheric measurements of magnetic and thermodynamic
  structure to estimate crudely the likely variation of Poynting flux
  across the surface immediately beneath the coronal loops.
  Measurements of chromospheric magnetic fields well above the
  $\beta=1$ surface avoids some of the difficulties associated with measurements
  of photospheric magnetic fields, but at the expense of 
  sacrificing any information on magnetic field components transverse to
  the line of sight. 
  }
In the present paper we study
four additional ViSP scans that were not reported in J24a, along with
a higher angular resolution scan in the infrared region from the GRIS
spectropolarimeter on the GREGOR telescope
(\citealp{Judge+others2024b}, henceforth ``J24b''). Both 
photospheric and chromospheric lines were observed by these 
instruments, building maps of surface conditions by scanning perpendicular to the slit. 
Consequently, velocity measurements, like the Zeeman measurements of chromospheric magnetic fields,  are restricted to components
along the line-of-sight.    
Some theoretical ideas are then pitted against these data in
section~\ref{sec:discussion}.

\figaodmm
\figawowp

\section{Poynting vectors and scalar proxies}
\subsection{Approximations}

In J24a and J24b, scalar ``proxies'' for Poynting vectors were
  used as first, crude estimates of the upward-directed Poynting flux
  emerging from the upper chromosphere into the corona.  In ideal MHD, the
  Poynting vector is
\begin{equation}
 \label{eq:poynting}
    \mathbf{S} = - \frac{c}{4\pi} (\mathbf{u}\times\mathbf{B}) \times
    \mathbf{B}.
\end{equation}
Projecting 
the fluid velocity vector
$\mathbf{u}$ into components 
perpendicular and parallel to
the magnetic field $\mathbf{B}$, which has unit vector $\hat{\mathbf{b}}$,
\begin{equation} \nonumber
    \mathbf{u}_\perp = (\hat{\mathbf{b}} \times \mathbf{u}) \times \hat{\mathbf{b}}; \ \ \ 
    \mathbf{u}_\parallel = \mathbf{u} -    \mathbf{u}_\perp     , 
\end{equation}
then (1) is simply
\begin{equation}
    \mathbf{S} = \frac{c}{4\pi} B^2\, u \, \hat{\mathbf{u}}_\perp.
    \label{eq:poyn}
\end{equation}
Clearly $\mathbf{S}$ is perpendicular to
$\mathbf{B}$, and $S= |\mathbf{S}|  \le
    \frac{c}{4\pi} B^2\, {u}.$
    
In comparison, our measurements of Doppler profiles and 
    Zeeman effect yield only  scalar quantities 
\begin{eqnarray} \label{eq:vlos}
    u_{LOS} &=& \mathbf{u} \cdot \hat{\mathbf{k}}, \ \ \ \ \mathrm{and}\\
    B_{LOS} &=& \mathbf{B} \cdot \hat{\mathbf{k}}, \label{eq:blos}
\end{eqnarray}
where unit vector $\hat{\mathbf{k}}$
lies along the line of sight. 
Therefore, based upon equation (2), in J24a, we compared
estimates of $S=|\mathbf{S}|$ with intensities of
the  coronal emission immediately above, using
 the crude scaling
\begin{equation}
    S \propto 
    B^2_{LOS}u_{LOS}, \label{eq:proxy}
\end{equation}
and a scaling appropriate for Alfv\'en waves (expression 7  below),
with $u_{LOS}$ measured from Doppler
widths and shifts of the cores of
chromospheric lines. 
Relation (5) assumes that some unknown but small fraction of the total magnetic energy flux (equation~\ref{eq:poyn}) is available for coronal heating.
 J24a showed that this 
quantity varied across the solar surface  similarly
to estimates using the scaling (7) for Alfv\'en 
waves.  Therefore, only the latter are shown in the present paper.

\subsection{Assumptions and limitations}
\label{subsec:limits}

The magnitude of 
$S$ depends on geometrical 
relations between vectors $\mathbf{B}$ and $\mathbf{u}$, which we cannot measure at present. 
So we face two choices:
first acknowledge that the observational
data are inadequate to make further
progress, and stop.  Second, make 
explicit 
assumptions:
\begin{enumerate}
    \item $u_{LOS}$ contains
information on $u_\perp$, and 
\item statistical correlations between chromospheric $\mathbf{u}$ and $\mathbf{B}$ are similar in the neighborhood of the footpoints of coronal plasma loops.
\end{enumerate}
The latter  is an hypothesis eventually to be tested, prompted by the observed universal presence of non-thermal,  unresolved motions in the chromosphere.

Equations (\ref{eq:poynting}) and 
(\ref{eq:poyn}) relate only solar 
properties, where vectors $\mathbf{u}$ and
$\mathbf{B}$ are independent.
Equations~(\ref{eq:vlos}) and (\ref{eq:blos})  
relate these two vectors to an entirely
different vector $\hat{\mathbf k}$
defined by the line of sight. 
For almost all lines of sight, it must be the case that the independent 
measurements of
$u_{LOS}$ and $B_{LOS}$  will always 
yield information about  $S$, except in three singular cases:  (a) $\mathbf{u}$ is parallel to $\mathbf{B}$, (b)
$\mathbf{u}$ is perpendicular to $\hat{\mathbf{k}}$, and 
(c) $\mathbf{B}$ is perpendicular to $\hat{\mathbf{k}}$.   Cases (b) and (c) are 
statistically rare, $\hat{\mathbf{k}}$ being entirely unrelated to the solar vectors.
Further, local verticals on the
  Sun are at least 0.33 radians from the observed LOS during all scans presented below 
  (heliographic coordinates are listed in the figure captions).  The
  local magnetic fields are statistically  significantly inclined to the
  LOS, i.e. at least the singular case $\mathbf{B}\cdot \hat{\mathbf{k}} = 0$ is a rare occurrence.

However, for strictly field-aligned
flows (case a), our analysis will lead to
spurious non-zero  estimates of
$S$. Such flows certainly
contribute 
to Doppler signals used below, potentially leading to  systematic errors in our analysis.  But two arguments 
suggest that
such errors are 
not overwhelming.
First, 
if 
our  assumption 2. is valid, they will
contribute only
a statistically constant offset in $S$ across 
the observed fields of view. Therefore we can 
partly justify  relating overlying coronal emission to systematic changes in $S$ across the field of view. Secondly such macroscopic flows along  fibrils 
are generally significantly
less than  unresolved motions
contributing to
widths of chromospheric lines 
\citep[e.g.][]{Athay1976}.

\figbpjdd

\subsection{Alfv\'en waves}

The remaining significant motions contributing to
$S$ are 
associated with 
Alfv\'en and fast modes, 
recalling that MHD turbulence 
is the non-linear interaction of Alfv\'en waves.
We estimate the Poynting flux as the product of
the wave group speed 
$c_A \sim B/\sqrt{4\pi\rho}$ with $B \sim B_{LOS}$ and wave energy densities.
The Alfv\'en 
waves carry an energy density 
along the ambient magnetic field
 $\mathbf{B}_0$
of 
\begin{equation}
    \varepsilon = \frac{B_1^2}{4\pi}= \rho u_1^2
\end{equation}
where $B_1$ and $u_1$ are the 
amplitudes of magnetic and velocity
perturbations perpendicular to $\mathbf{B}_0$ \citep{Braginskii1965}.   Velocity
perturbations $u_1^2$
can be estimated from the non-thermal widths
of chromospheric lines, which tend to exceed 
statistical fluctuations in 
Doppler shifts of line  centroids.
We  therefore examine the 
proxy \citep{Jordan+Mendoza+Gill1984,
Jordan+others1987}:
\begin{equation} \label{eq:waves}
    S \propto \frac{B_{LOS}}{\sqrt{4\pi\rho}} \, \rho\,  \xi^2,
\end{equation}
using non-thermal spectral line widths $\xi$ to estimate $u_1$.  The densities of the 
plasma at the formation heights
of the \ion{Ca}{2} and \ion{He}{1}
line cores are given in J24a,b.   Many observations show that $\xi$ 
in chromospheric lines statistically increases towards the solar limb
\citep[][see their page 180] {Linsky+Avrett1970}, 
a behavior compatible with wave motions transverse to 
statistically radial magnetic fields within coronal loop footpoints. 

\figbnkvm

\section{Recent measurements of chromospheric
  magnetic fields}

 We analyze data from two ground-based observatories to complement
the analysis of J24a, which was concerned with just one scan of ViSP
chromospheric magnetism in relation to data from the HMI
\citep{Hoeksema+others2018} and AIA instruments on SDO, and the IRIS
UV instrument \citep{IRIS} . Here we examine chromospheric magnetic fields measured
closer still to the base of the corona, by studying the 1083 nm line
of \ion{He}{1}, and also analyze data from the four other scans in
\ion{Ca}{2} obtained during the series of observations reported by
J24a.

{The \ion{Ca}{2} line core forms just $\sim2$ pressure scale
  heights beneath the corona \citep{Cauzzi+others2008,
    Leenaarts+others2009}, 8-9 pressure scale heights above the
  photosphere.   With lower limits to
  field strengths measured here of $\sim 200$ G, the plasma $\beta=1$ surface 
will be $\lesssim$ 400 km, $\lesssim3$ pressure scale heights, above the photosphere.

  The lines of the \ion{He}{1} 1083 nm multiplet core
  form in tenuous and dynamic plasma close to the base of the corona,
  mostly above the stratified chromosphere, where ionizing radiation
  can excite levels within the triplet system of helium
  \citep{Leenaarts+others2016}.}

\subsection{GRIS data from 28 September 2020}

{J24b analyzed data from the GRIS instrument
  \citep{Collados+others2012} on the GREGOR telescope
  \citep{GREGOR2012,Kleint+others2020}, but using the lines of
  \ion{Si}{1} 1082.7 and \ion{He}{1} 1083 nm from the GREGOR
  telescope. } These infrared data were acquired by Tobias Felipe and
Christoph Kuckein on 28 Sep 2020, {they are discussed in detail by
  J24b. } Figure~\ref{fig:gris} shows magnetic fields derived from the
GRIS scan in lines of \ion{Si}{1} at 1082.7 and \ion{He}{1} at 1083.0
nm.  The \ion{Si}{1} line forms at 200-500 km above the continuum
photosphere \citep{Judge+others2014}.
  Using the measured non-thermal 
  linewidth $\xi$ in place of $u_{LOS}$
in equation~(\ref{eq:waves}), 
$S \propto \rho\xi^2c_A$ is shown in the rightmost panel of
Fig.~\ref{fig:gris}, and in the lowest rows of all other figures.  
For
reference P24b used $\rho \sim 10^{-13}$ g~cm$^{-3}$ as an estimate of
the plasma mass density at the formation height of \ion{He}{1} 1083 nm
\citep{Avrett+Fontenla+Loeser1994}. In this and later figures the
heliographic solar $X$ and $Y$ coordinates are given. The angle between lines of sight and
local vertical on the Sun is 0.33 radians for
Figure~\ref{fig:gris}, and up to 0.74 radians for the ViSP scans in later figures. Only
\textit{differences} between these estimates across the solar surface
can be assumed to be meaningful.

The lowest panel of Figure~\ref{fig:gris} is morphologically
  similar to a figure, not shown here, of the other Poynting flux
  scalar.  Both figures demonstrate the fact that
\textit{no correlation is found between coronal plasma loop footpoints and
  underlying chromospheric $B_{LOS}$ magnetic fields and proxies for
  Poynting flux}.  For example, the yellow circle
identifies a loop footpoint, the blue circle a region of stronger
magnetic field and Poynting flux with no associated loop footpoint.
These are just two of many examples. 

This lack of correlation does not depend on projection effects or the
small uncertainties ($\sim \pm2\arcsec$, J24b) in co- alignment.
The GRIS data are for an active region which is inclined at $\approx 0.33$ radians to the LOS. The
difference in formation height between 1800 km near where the
\ion{Ca}{2} and \ion{He}{1} line cores are formed, and the onset of
17.1 nm mission is $\approx1$ Mm in conductively coupled models
\citep[e.g.][]{Hansteen1993}. Therefore a vertical plasma loop would
appear displaced by $\pm\sim 0.7$ Mm ($< 1$ \arcsec) from the
chromosphere along the (unknown) tube's field lines.  This offset is smaller
than the estimated co-alignment accuracy.  However, 
photospheric images would be displaced in projection by $ \sim2.5$ Mm.

\subsection{DKIST data from 02-03 June 2022}

J24a focused on just one scan with the ViSP instrument obtained on 3
June 2022.  Data of the 630 nm \ion{Fe}{1} line pair and the 854.2 nm
chromospheric line of \ion{Ca}{2} were used to extract properties of
the plasma and magnetic field.  Four similar ViSP scans were made
between June 2 2022 20:02 and June 3 18:50 UT (see Table 1 of J24a).
All data were analyzed with co-aligned supporting data from
instruments on the SDO spacecraft.  In Figures~\ref{fig:aodmm} -
\ref{fig:bnkvm}, data from the remaining four scans are shown in a
similar manner to Figure~\ref{fig:brwjv}.  {Measured LOS
  chromospheric magnetic flux densities over the photospheric flux
  concentrations} are typically 250 Mx~cm$^{-2}$ (J24a,J24b).  Clearly
the behavior found by J24a is \textit{typical of all these cases},
uncovered through measurements of lines of \ion{Fe}{1} at 630 nm and
\ion{Ca}{2} at 854 nm, formed at roughly 250 and 1500 km above the
continuum photosphere respectively (\citealp{Grec+others2010,
  Cauzzi+others2008}).

\section{Discussion}
\label{sec:discussion}

Six separate scans show that {coronal} plasma loop footpoints
connect to a small area of broader chromospheric magnetic fluxes
beneath.  All scans contain 5-6 plasma loop footpoints.  Of these
  $\approx 30$ footpoints, the probability that we missed strong
  corrrelations of loops with enhanced Poynting fluxes all at the
  ``other'' (unobserved) footpoint is negligible ($\sim 0.5^{30}$).
The footpoints do tend to lie close to the edges of patches of
chromospheric magnetic fields.  The same is not true for the plasmas
emitting transition region radiation (J24a documented this
observation, which included transition region data from IRIS as
  well as SDO in section 3.3 and their Appendix).     These
statistically robust results present new and essential
  observational information concerning conditions at the footpoints of
  typical active region coronal loops. 

While 
  secular
  motions $u_\perp$ driving magnetic fields send  energy into the corona \citep[e.g.][]{Parker1988}, higher frequency contributions to $S$ 
  can be reduced. In a homogeneous 1D model, $\gtrsim 50\%$ of the wave power can be reflected
  analogously to an open boundary on an oscillating string, because of 
  the steep density gradient at the 1D transition region \citep{Wentzel1978}. But less coherent waves 
  in an inhomogeneous medium 
  can intermingle, become non-linear, leading to 
  a richer variety of phenomena
(\citealp{Wentzel1978}, a modern
perspective is summarized by \citealp{Cranmer+Molnar2023}).  In this paper we have assumed 
  that statistically, a constant fraction of any Alfv\'en wave energy emerges through
  the transition region,  no matter how complex the dynamics.  This assumption 
  is questionable. But we note that the structure of transition region plasma remains poorly
  understood (see J24a,b, also
  \citealp{Judge+Ionson2024}), and that
  the overlying corona and transition region are
  \textit{products} of the unknown energy transport
  emerging from the chromosphere. Thus their structure may well differ from a
  prescribed solution to a steady, 1D energy equation, previously used to analyze  reflection coefficients.

\subsection{The new observations in context}

In spite of trade-offs noted above, the direct physical connection of
the chromosphere to the corona is but one reason that chromospheric
data are superior to photospheric measurements, for connecting lower
to upper atmospheric structures.  Another is that {\it observations
  are superior to {magnetic field} extrapolations}, particularly
given that between photosphere and chromosphere lies the awkward
transition from fluid- to magnetic- dominated stresses near the
$\beta=1$ surface. 

All six scans of chomospheric magnetic fields support one elementary
fact: \textit{something missing from current observational
  capabilities must be an important ingredient in the way the corona
  is heated. }  Only a small fraction of the
area covered by $|B| \gtrsim 200$ Mx~cm$^{-2}$ unipolar chromospheric
fields correspond to coronal loop footpoints.

LW93 recognized this problem during an era when only photospheric
magnetic fields were routinely measured, at significantly lower
angular resolutions:
\begin{quote}
    ``Why are coronal loops ``rare", in the sense that only a small
  subset of coronal magnetic field lines become loaded with hot
  plasma?\ldots The relative rarity of coronal loops implies that,
  whatever the heating mechanism, it must have a relatively high
  threshold so that conditions for its onset are satisfied only in a
  small fraction of the available coronal volume.''
\end{quote}
The existence of a high threshold for the ``turn-on'' of coronal
heating appears at odds with much work on the thermal physics of
coronal loops cited earlier.  Field-aligned, one dimensional
hydrodynamic calculations driven by various kinds of ad-hoc heating
parameters, and subject to different formulations for radiation
losses, tend to exhibit simple evolution towards {unique}
equilibrium states. Only under special prescribed conditions, namely enhanced heating rates close to footpoints, can exhibit 
 limit-cycle behavior \citep[e.g.][]{Mueller+others2003, Martens2010}
as first found in fieldline-integrated thermodynamic calculations of \citet{Kuin+Martens1982}.
  One might also postulate that time-dependent driving of the loop system
  naturally leads to diverse observed loop properties
  \citep[e.g.][]{Hansteen1993}.  It remains to be seen if, statistically,  the observed  independence
  of footpoint brightness from estimates of Poynting flux is consistent with this kind of  interpretation.

The present analysis should be understood in terms of the response of
the solar upper atmosphere to forcing from a convectively-dominated
atmosphere below.  In this sense it differs from analyses of data
associated with ``plumes'' or loops originating in the umbrae of
sunspots \citep{Foukal1975,Foukal1976,Chitta+others2016}, for which
convection is suppressed.  Our analysis is therefore more applicable
to the more general heating problem than special cases for sunspot
umbrae \citep[e.g.][]{Foukal1975}, which \citet{Ionson1978}
modeled in terms of the 
resonant absorption picture, which 
continues to be studied via numerical
experimentation
\citep[e.g.][]{Belien+others1999,Howson+others2020}.

\subsection{Locally multipolar fields, {reconnection} and plasma heating}

The tendency for loop footpoints to appear towards the edges of
chromospheric network naturally suggests that reconnection with
minority polarity internetwork fields might supply energy for the
heating of these loops \citep{Title+Schrijver1998,
  Priest+others2002,Wang2016,Chitta+others2017,Chitta+others2023}.
{Curiously, several inferences from the observations of the
  chromosphere under the loop footpoints suggest otherwise.  In the
  Appendix we argue that magnetic fields under the observed plasma
  loops must be unipolar, supporting only small-angle discontinuous
  fields, in order to satisfy observational data along with force
  balance. }

The plasma $\beta$ is $\lesssim 0.002$ in the topmost scale height of
the chromosphere, using $B_0\gtrsim 250 $G from our line-of-sight
measurements and a plasma pressure of 4 dyne~cm$^{-2}$ from model F of
\citet{Vernazza+Avrett+Loeser1981}.  Consequently the intrinsic
magnetic field strength $B_0$ must be nearly constant there, in order
to maintain horizontal pressure balance.  {Along with the
  relatively uniform measurements of $B_{LOS}$, this condition leads
  to a contradiction, denying a role for such reconnection at the foot
  of the plasma loops.}

\subsection{One Poynting flux, two outcomes?}

\newcommand{\poynt}{{\ensuremath{\mathbf{S}}}}

The two proxies used both by J24a and J24b for the scalar amplitude of the Poynting flux $S$
are far from satisfactory. Not even the sign of the vertical flux can
be inferred.
The chromospheric network characterized by $B_{LOS} \gtrsim 200$
Mx~cm$^{-2}$ occupies several thousand pixels in each scan. Therefore
local averages can be used to compare relative sizes of the proxies
between regions near plasma loop
footpoints and other regions.  It is only in this sense that we might
claim that $S$ does not change much across the bright
network boundaries.  If we accept this rationale, then all of the
figures show that something not encoded in the data presented has
primary control of the overlying corona (J24a), of overlying loops
would have emerged.  for otherwise some relation between
$S$ and the brightness overlying loops would have emerged.
Theoretically, the question of what might be responsible for the
behavior has followed two trains of thought, discussed next: First, both static and hydro-dynamic thermal
models with parameterized heating rates have been studied for the
presence of multiple long-lived solutions.
Second, 
instabilities of dissipative structures, be they of MHD or
electrostatic origin (LW93), have been studied.

\subsubsection{Thermal models}

Can purely thermal models lead to different long-lived states of
coronal loops for the same dissipated Poynting flux?  By ``thermal''
we refer to single fluid, field-aligned hydrodynamics energy transport
along a one dimensional representation of a plasma loop
\citep{Boris+Mariska1982,Mariska+others1982, Hansteen1993,
  Bradshaw+Mason2003,Bradshaw+Cargill2013}.  On surveying the
literature, the consensus seems to be ``no'', except under certain
special conditions \citep{Muller+Hansteen+Peter2003,Martens2010}.
``Zero-dimensional'' studies in which the energy equation is
integrated analytically along the loop have also been useful,
revealing two kinds of long-lived structures, although not without
some initial confusion \citep{Kuin+Martens1982,Craig+Schulkes1985}.
More recent formulations have resolved the differences in favor of
evolution towards single equilibria
\citep{Vandenoord+Barstow1988,EBTEL, Cargill+Bradshaw+Klimchuk2012a,
  Cargill+Bradshaw+Klimchuk2012b}.  Following
\citet[][]{Rosner+Tucker+Vaiana1978} most such calculations adopt
parameterized forms of volumetric heating functions, typically
\begin{equation}\label{eq:heat}
    \dot E \sim T^a n^b f(s)
\end{equation}
where $T$ and $n$ are plasma temperature and density, respectively,
and $s$ distance along the loop.  With ``reasonable'' values of $|a|$
and $|b|$ less than, say, 2-3, the solutions are unique for given
rates of heating $\dot E$.  One stabilizing process is a strong
exchange of mass and energy between the stratified chromosphere and
hot coronal plasma, driven by downward heat conduction.  Another stabilizer
occurs when the heating $f(s)$ as a function of distance $s$ along the loop
does not fall off too dramatically with height. If it does, then
cooling ensues and the system evolves dynamically as dynamic
equilibrium is lost.  Occasionally, specific formulations exhibit
limit cycle behavior. Depending on initial states, a limit cycle
solution and a different equilibrium solution might explain
qualitatively our main results.  The ``0D'' model
\citet{Kuin+Martens1982} exhibited such behavior, and tests revealed
that the two kinds of solutions result from insufficient coupling
between chromosphere and corona (see also
\citealp{Vandenoord+Barstow1988}).  \citet{Muller+Hansteen+Peter2003}
found limit cycle behavior in their 1D calculations of a short plasma
loop, depending on the damping length $k$ using a heating function
$f(s) = \exp(-s/\ell$).

Firmer conclusions are difficult to draw from these thermal models, in
part because of the several ad-hoc assumptions behind them and the
consequent large parameter space to be explored.  Prescriptions based
upon equation (\ref{eq:heat}) amount to a ``closure'' of simplified
equations of motion, determining all outcomes including those relevant
for stability. They have no justification in terms of physical
dissipation mechanisms within real plasmas.  At this stage we
therefore regard the question posed above as unanswered.

\subsubsection{MHD and plasma processes}

Here we follow the suggestion of LW93, and consider that a solution
may be found among the physical heating mechanisms themselves.  The
authors, noting that almost all electrodynamic models of irreversible
dissipative processes operate on unobservably small scales,
nevertheless concluded that such models could explain neither the
observed intermittency nor the apparent transverse physical scale
across plasma loops observable at the time.

Of all the irreversible processes discussed in section 3 of LW93, only
their model of reconnection between adjacent, tangentially mis-aligned
models might satisfy observed constraints. In particular the
transverse thicknesses of plasma loops seemed to require interactions
of elemental flux tubes (identified as such in the photosphere),
essentially Parker's (1988) picture of nanoflare heating. However, the
lack of discrete tubes seen in modern observations of the solar
photosphere suggests that elemental structures may consist of tortuous
eddies of magnetic flux formed by convective dynamics
\citep{Rast+others2021}.  Such ideas are worth pursuing, especially
with the new observational capabilities offered by DKIST.
 
The EUI instrument \citep{EUI} on the Solar Orbiter mission
\citep{Orbiter,Orbiter2}, currently achieves a similar resolution to
that found by \citet{Rachmeler+2022}, close to 240 km.  Further
detailed studies of EUI data should thus be a priority, to explore
these two pictures (ion viscosity versus topological change due to
reconnection).  A scale of 240 km is close to that at which ion-ion
collisions (``viscosity'') can thermalize compressive motions
\citep{Hollweg1986,Davila1987,Judge+Ionson2024}, driven
by a variety of processes which might include MHD turbulence, wave
motions in inhomogeneous conditions and reconnnection.  Future studies
might be able to rule out the viscous picture if scales well below,
say, 200 km can be found within coronal loops, since then dissipation
must occur on scales below  the current observational horizon
\citep{Judge+Ionson2024}.

Intriguing behavior has been found in 
analytical \citep{Uzdensky2007} 
and numerical \citep{Einaudi+others2021} studies of non linear plasma
dynamics.   \citet{Uzdensky2007} appeals to 
a transition from slow to fast magnetic 
reconnection regimes according to whether 
coronal loop plasmas are collisional or
collisionless systems. \citet{Einaudi+others2021}
found in their MHD numerical turbulence experiment
characteristics of a \textit{self-regulating system,} in which the
``state [of the plasma] occurs independently of the detailed form of
the boundary velocity.''  The driving energy used in their idealized
``plasma loop'' was about $3 \times 10^5$ erg~cm~$^{-2}$~s$^{-1}$,
about 30 times smaller that requirements for active region plasma.
They concluded that non-linear processes occurring within the corona
itself determine the spatial and temporal properties within the fluid.
Perhaps such effects, or similar non-linearities are necessary to
organize coronal plasma loops into macroscopic states which resemble
the intermittently, but not randomly, ordered corona.

\section{Conclusions}

Coronal heating is and must always be an observationally-led research
area.  In the near future we anticipate the advent of further new
idealized numerical experiments, novel data from observational
facilities such as GREGOR and DKIST, and from instruments such as
those on the Solar Orbiter spacecraft, which will help illuminate the
problems raised in this study. Future observational work
should attempt to understand the nature of
unresolved motions in the upper chromosphere
in relation to the long-lived magnetic fields there. 
POS motions and, if possible, POS chromospheric magnetic field components should also be included
to study this problem.
Measurements of magnetic fields from photosphere
to coronal base should, assuming magnetic field line 
continuity, help resolve the 
ambiguity in the POS chromospheric 
magnetic field vector, by tracing fields from
their single polarity roots into canopy-like 
configurations.

Given our assumptions, data and analysis, we are faced with a potential conundrum: What happens where there is clearly a
signature of chromospheric Poynting flux but no coronal loop
footpoints? Is there an unseen corona between the visible coronal
loops?  If indeed the chromospheric Poynting flux estimates with no
  associated visible coronal loop inject energy into the corona, where
  does it go? Does it escape to infinity or become stored in the wider
  corona? Could the coronal heating be a secondary result of such processes?

The observations suggests that the ``empty" corona between
loops may be in a different, yet long-lived state. Otherwise we might
argue that those ``empty" loops would fill up with plasma, much like
\citet{Kuin+Martens1982} envisioned in their original model. Perhaps
there is some sort of threshold for forming the hot coronal loops, as
suggested by LW93.  {The high threshold for onset of coronal
  heating above active regions (LW93) finds a natural explanation in
  the switch from slow fluid reconnection to fast collisionless
  reconnection, proposed by \citet{Uzdensky2007}.  Future work on this
  proposal seems worthwhile.  Scalings to conditions between main
  sequence and evolved stars do not discount this picture (work in
  progress).  }

As a final thought, this work reminds us to seek elementary and robust
empirical facts about remotely sensed astronomical objects, before we
throw the full machinery of numerical exploration, machine learning
and advanced, detailed modeling at certain problems \citep[see
  also][]{Judge+Ionson2024}.

\section*{Acknowledgments}

The authors thanks Lucia Kleint and the astronomy department at the
University of Bern and the Swiss National Science Foundation (grant
No. 216870) which made this work possible.  This material is based
upon work supported by the National Center for Atmospheric Research,
which is a major facility sponsored by the National Science Foundation
under Cooperative Agreement No. 1852977.  Comments from Peter Cargill and a patient anonymous referee
greatly influenced this article, and are gratefully acknowledged.

\appendix
\section{Multipolar chromospheric magnetic fields?}

{In this appendix we extend conceptual models of reconnection of
  multipolar photospheric fields as a driver of coronal heating of
  plasma loops, up into the chromosphere where our critical
  observations are made.  Thus we give the models proposed by
  \citet{Title+Schrijver1998,Priest+others2002,Wang2016,Chitta+others2017,Chitta+others2023}.  the benefit of the doubt,
  to see if they remain viable when confronted with chromospheric
  field measurements.  }

{Unlike the photosphere, the measurements of chromospheric fields
  are from regions of $\beta \ll 1$, thus the field strengths
  $|\mathbf{B}|$ in such regions must locally be almost identical, independent
  of the direction of the field line vectors.  Let us consider then
  the case where each observed pixel receives photons from a
  distribution of orientations of magnetic field vectors all with the
  same strength $B_{0}$. Then the magnetic field inferred from the
  circularly polarized line profile (Stokes $V$) relative to intensity $I$ yields, in
  each pixel, }
\begin{equation}
B_{LOS}= \langle B_0\cos\theta \rangle= \frac{B_0}{A} \int_A
\cos\theta(x,y) dx dy. \label{eq:long}
\end{equation}
{Here $x$ and $y$ specify geometric coordinates in the plane
  perpendicular to the LOS, and $A$ is the projected area of one
  detector pixel onto this plane.  $0 \le \theta \le \pi$ are the
  (also unknown) angles of inclination of the fields to the line of
  sight ($\theta=0$ being parallel to it, $\theta=\pi$ anti-parallel).
  In the text we find $B_{LOS}$ is typically $250$ Mx~cm$^{-2}$, and
  is of the same sign beneath and in the neighborhood of 
  each observed loop footpoint. }

{In the above equation, $B_0$ lies outside of the integral owing
  to the smallness of $\beta$.  To illuminate implications  of
  unresolved chromospheric magnetic fields of opposite polarity, we
  can assume that $\cos\theta$ under the integral has just two
  components, parallel and antiparallel to the LOS, which reconnect,
  releasing magnetic free energy for subsequent heating. Then}
\begin{equation}
B_{LOS}= {B_0} (1-f -f) = {B_0}(1-2f)
\label{eq:longapp}
\end{equation}
{where $f < 0.5$ is the fractional area of the minority
  component. When $f \ll 1$, the measured $B_{LOS} \rightarrow
  B_0$.  In the other limit where $f = 1/2 -\delta$, 
  $\delta \ll 1$,  $B_{LOS}
  \rightarrow B_0\delta  \ll B_0$. Next let us assume, as argued by those
  advocating for multipolar fields as agents of heating, that each
  loop footpoint sits above a case where $f$ is finite, and that
  outside each footpoint $f$ is negligibly small.  The basic
  observational fact from our analysis in the text is that $B_{LOS}$
  is roughly constant, say $B_1$, across the loop footpoint and its
  immediate surroundings.  A clear example is shown in in Figure~6 of
  J24a, in the neighborhood of the bright EUV footpoint at $X=-403$,
  $Y=-385$.  Within those pixels with mixed LOS polarity we have
  $B_{LOS,mixed} = B_1 = B_0 \delta$.  Outside this region where
  $\delta$ is negligible, we have $B_{LOS,unipolar}=B_1 = B_0$.  The
  data show that $B_{LOS,mixed} \approx B_{LOS,unipolar}$, so that we
  arrive at $B_0 = B_0\delta$, which is a contradiction since $0 \le
  \delta \le 1/2$.  }

{This argument is supported by noting that any reconnection
  associated with the annihilation of opposite polarity fields} should
be accompanied by Doppler signatures of vertical motions $ u \lesssim
c_A \approx 1000$ km~s$^{-1}$ in the upper chromosphere.  Yet the
movie (Figure~6 of J24b) of H$\alpha$ core intensities shows no
indication of changes of magnetic connectivity in the upper
chromosphere during the duration of the GRIS observations.  Further,
observed {spectral linewidths and shifts of chromospheric and
  coronal lines } are typically two orders of magnitude smaller.
{(see Appendix~A of \citealp{Judge+Ionson2024}).}

We conclude that it does not seem possible to heat the loops
associated with unipolar regions by invoking the reconnection of 
minority polarity fields lying below current limits of observability,
without violating some basic observational constraints.  Instead,
solutions must be sought based upon locally unipolar fields
\citep{Judge2021}.

\bibliographystyle{aasjournal}

\begin{thebibliography}{}
\expandafter\ifx\csname natexlab\endcsname\relax\def\natexlab#1{#1}\fi
\providecommand{\url}[1]{\href{#1}{#1}}
\providecommand{\dodoi}[1]{doi:~\href{http://doi.org/#1}{\nolinkurl{#1}}}
\providecommand{\doeprint}[1]{\href{http://ascl.net/#1}{\nolinkurl{http://ascl.net/#1}}}
\providecommand{\doarXiv}[1]{\href{https://arxiv.org/abs/#1}{\nolinkurl{https://arxiv.org/abs/#1}}}

\bibitem[{{Aschwanden} {et~al.}(2000){Aschwanden}, {Nightingale}, \& {Alexander}}]{Aschwanden+others2000}
{Aschwanden}, M.~J., {Nightingale}, R.~W., \& {Alexander}, D. 2000, ApJ, 541, 1059

\bibitem[{{Aschwanden} {et~al.}(2007){Aschwanden}, {Winebarger}, {Tsiklauri}, \& {Peter}}]{Aschwanden+others2007}
{Aschwanden}, M.~J., {Winebarger}, A., {Tsiklauri}, D., \& {Peter}, H. 2007, ApJ, 659, 1673

\bibitem[{Athay(1976)}]{Athay1976}
Athay, R.~G. 1976, The Solar Chromosphere and Corona: Quiet Sun (Dordrecht: Reidel)

\bibitem[{Avrett {et~al.}(1994)Avrett, Fontenla, \& Loeser}]{Avrett+Fontenla+Loeser1994}
Avrett, E.~H., Fontenla, J.~M., \& Loeser, R. 1994, in Infrared Solar Physics, ed. D.~M. Rabin, J.~T. Jefferies, \& C.~Lindsey, Proc.\ IAU Symp.\ 154 (Dordrecht: Kluwer), 35--47t

\bibitem[{{Barczynski} {et~al.}(2018){Barczynski}, {Peter}, {Chitta}, \& {Solanki}}]{Barczynski+others2018}
{Barczynski}, K., {Peter}, H., {Chitta}, L.~P., \& {Solanki}, S.~K. 2018, A\&A, 619, A5, \dodoi{10.1051/0004-6361/201731650}

\bibitem[{{Beli{\"e}n} {et~al.}(1999){Beli{\"e}n}, {Martens}, \& {Keppens}}]{Belien+others1999}
{Beli{\"e}n}, A.~J.~C., {Martens}, P.~C.~H., \& {Keppens}, R. 1999, ApJ, 526, 478, \dodoi{10.1086/307980}

\bibitem[{{Boris} \& {Mariska}(1982)}]{Boris+Mariska1982}
{Boris}, J.~P., \& {Mariska}, J.~T. 1982, ApJL, 258, L49, \dodoi{10.1086/183828}

\bibitem[{{Bradshaw} \& {Cargill}(2013)}]{Bradshaw+Cargill2013}
{Bradshaw}, S.~J., \& {Cargill}, P.~J. 2013, ApJ, 770, 12, \dodoi{10.1088/0004-637X/770/1/12}

\bibitem[{{Bradshaw} \& {Mason}(2003)}]{Bradshaw+Mason2003}
{Bradshaw}, S.~J., \& {Mason}, H.~E. 2003, A\&A, 401, 699, \dodoi{10.1051/0004-6361:20030089}

\bibitem[{Braginskii(1965)}]{Braginskii1965}
Braginskii, S.~I. 1965, Reviews of Plasma Physics., 1, 205

\bibitem[{{Cargill} {et~al.}(2012{\natexlab{a}}){Cargill}, {Bradshaw}, \& {Klimchuk}}]{Cargill+Bradshaw+Klimchuk2012b}
{Cargill}, P.~J., {Bradshaw}, S.~J., \& {Klimchuk}, J.~A. 2012{\natexlab{a}}, ApJ, 758, 5, \dodoi{10.1088/0004-637X/758/1/5}

\bibitem[{{Cargill} {et~al.}(2012{\natexlab{b}}){Cargill}, {Bradshaw}, \& {Klimchuk}}]{Cargill+Bradshaw+Klimchuk2012a}
---. 2012{\natexlab{b}}, ApJ, 752, 161, \dodoi{10.1088/0004-637X/752/2/161}

\bibitem[{{Cauzzi} {et~al.}(2008){Cauzzi}, {Reardon}, {Uitenbroek}, {Cavallini}, {Falchi}, {et~al.}}]{Cauzzi+others2008}
{Cauzzi}, G., {Reardon}, K.~P., {Uitenbroek}, H., {et~al.} 2008, A\&A, 480, 515

\bibitem[{{Chitta} {et~al.}(2016){Chitta}, {Peter}, \& {Young}}]{Chitta+others2016}
{Chitta}, L.~P., {Peter}, H., \& {Young}, P.~R. 2016, A\&A, 587, A20, \dodoi{10.1051/0004-6361/201527340}

\bibitem[{{Chitta} {et~al.}(2017){Chitta}, {Peter}, {Solanki}, {Barthol}, {Gandorfer}, {Gizon}, {Hirzberger}, {Riethm{\"u}ller}, {van Noort}, {Blanco Rodr{\'\i}guez}, {Del Toro Iniesta}, {Orozco Su{\'a}rez}, {Schmidt}, {Mart{\'\i}nez Pillet}, \& {Kn{\"o}lker}}]{Chitta+others2017}
{Chitta}, L.~P., {Peter}, H., {Solanki}, S.~K., {et~al.} 2017, ApJS, 229, 4, \dodoi{10.3847/1538-4365/229/1/4}

\bibitem[{{Chitta} {et~al.}(2023){Chitta}, {Solanki}, {del Toro Iniesta}, {Woch}, {Calchetti}, {Gandorfer}, {Hirzberger}, {Kahil}, {Valori}, {Orozco Su{\'a}rez}, {Strecker}, {Appourchaux}, {Volkmer}, {Peter}, {Mandal}, {Aznar Cuadrado}, {Teriaca}, {Sch{\"u}hle}, {Berghmans}, {Verbeeck}, {Zhukov}, \& {Priest}}]{Chitta+others2023}
{Chitta}, L.~P., {Solanki}, S.~K., {del Toro Iniesta}, J.~C., {et~al.} 2023, ApJL, 956, L1, \dodoi{10.3847/2041-8213/acf136}

\bibitem[{{Collados} {et~al.}(2012){Collados}, {L{\'o}pez}, {P{\'a}ez}, {Hern{\'a}ndez}, {Reyes}, {Calcines}, {Ballesteros}, {D{\'\i}az}, {Denker}, {Lagg}, {Schlichenmaier}, {Schmidt}, {Solanki}, {Strassmeier}, {von der L{\"u}he}, \& {Volkmer}}]{Collados+others2012}
{Collados}, M., {L{\'o}pez}, R., {P{\'a}ez}, E., {et~al.} 2012, Astronomische Nachrichten, 333, 872, \dodoi{10.1002/asna.201211738}

\bibitem[{{Craig}(1990)}]{Craig1990}
{Craig}, I.~J.~D. 1990, A\&A, 234, L12

\bibitem[{{Craig} \& {Schulkes}(1985)}]{Craig+Schulkes1985}
{Craig}, I.~J.~D., \& {Schulkes}, R.~M.~S.~M. 1985, ApJ, 296, 710, \dodoi{10.1086/163488}

\bibitem[{{Cranmer} \& {Molnar}(2023)}]{Cranmer+Molnar2023}
{Cranmer}, S.~R., \& {Molnar}, M.~E. 2023, ApJ, 955, 68, \dodoi{10.3847/1538-4357/acee6c}

\bibitem[{Davila(1987)}]{Davila1987}
Davila, J.~M. 1987, ApJ, 317, 514

\bibitem[{{de Wijn} {et~al.}(2022){de Wijn}, {Casini}, {Carlile}, {Lecinski}, {Sewell}, {et~al.}}]{Dewijn+others2022}
{de Wijn}, A.~G., {Casini}, R., {Carlile}, A., {et~al.} 2022, Solar Phys., 297, 22

\bibitem[{{Einaudi} {et~al.}(2021){Einaudi}, {Dahlburg}, {Ugarte-Urra}, {Reep}, {Rappazzo}, \& {Velli}}]{Einaudi+others2021}
{Einaudi}, G., {Dahlburg}, R.~B., {Ugarte-Urra}, I., {et~al.} 2021, ApJ, 910, 84

\bibitem[{{Foukal}(1975)}]{Foukal1975}
{Foukal}, P. 1975, Sol. Phys., 43, 327

\bibitem[{Foukal(1976)}]{Foukal1976}
Foukal, P. 1976, ApJ, 210, 575

\bibitem[{{Golub} {et~al.}(1980){Golub}, {Maxson}, {Rosner}, {Vaiana}, \& {Serio}}]{Golub+others1980}
{Golub}, L., {Maxson}, C., {Rosner}, R., {Vaiana}, G.~S., \& {Serio}, S. 1980, ApJ, 238, 343, \dodoi{10.1086/157990}

\bibitem[{{Grec} {et~al.}(2010){Grec}, {Uitenbroek}, {Faurobert}, \& {Aime}}]{Grec+others2010}
{Grec}, C., {Uitenbroek}, H., {Faurobert}, M., \& {Aime}, C. 2010, A\&A, 514, A91, \dodoi{10.1051/0004-6361/200811455}

\bibitem[{Hansteen(1993)}]{Hansteen1993}
Hansteen, V. 1993, ApJ, 402, 741

\bibitem[{{Hoeksema} {et~al.}(2018){Hoeksema}, {Baldner}, {Bush}, {Schou}, \& {Scherrer}}]{Hoeksema+others2018}
{Hoeksema}, J.~T., {Baldner}, C.~S., {Bush}, R.~I., {Schou}, J., \& {Scherrer}, P.~H. 2018, Sol. Phys, 293, 45, \dodoi{10.1007/s11207-018-1259-8}

\bibitem[{Hollweg(1986)}]{Hollweg1986}
Hollweg, J.~V. 1986, ApJ, 306, 730

\bibitem[{{Howson} {et~al.}(2020){Howson}, {de Moortel}, \& {Reid}}]{Howson+others2020}
{Howson}, T.~A., {de Moortel}, I., \& {Reid}, J. 2020, A\&A, 636, A40

\bibitem[{{Ionson}(1978)}]{Ionson1978}
{Ionson}, J.~A. 1978, ApJ, 226, 650

\bibitem[{{Jordan} {et~al.}(1987){Jordan}, {Ayres}, {Brown}, {Linsky}, \& {Simon}}]{Jordan+others1987}
{Jordan}, C., {Ayres}, T.~R., {Brown}, A., {Linsky}, J.~L., \& {Simon}, T. 1987, MNRAS, 225, 903, \dodoi{10.1093/mnras/225.4.903}

\bibitem[{{Jordan} {et~al.}(1984){Jordan}, {Mendoza}, \& {Gill}}]{Jordan+Mendoza+Gill1984}
{Jordan}, C., {Mendoza}, B., \& {Gill}, R.~S. 1984, in ESA-SP 220, ed. T.~D. {Guyenne} \& J.~J. {Hunt} (ESA), 133--136

\bibitem[{{Judge} {et~al.}(2024{\natexlab{a}}){Judge}, Kleint, {Casini}, de~Wijn, {Schad}, \& {Tritschler}}]{Judge+others2024a}
{Judge}, P., Kleint, L., {Casini}, R., {et~al.} 2024{\natexlab{a}}, ApJ, 960, 129

\bibitem[{{Judge} {et~al.}(2024{\natexlab{b}}){Judge}, Kleint, \& {Kuckein}}]{Judge+others2024b}
{Judge}, P., Kleint, L., \& {Kuckein}, C. 2024{\natexlab{b}}, ApJ submitted, February 2024

\bibitem[{{Judge}(2021)}]{Judge2021}
{Judge}, P.~G. 2021, ApJ, 914, 70

\bibitem[{{Judge} \& {Ionson}(2024)}]{Judge+Ionson2024}
{Judge}, P.~G., \& {Ionson}, J.~A. 2024, The Problem of Coronal Heating. A Rosetta Stone for Electrodynamic Coupling in Cosmic Plasma (Springer, in press, March 2024)

\bibitem[{{Judge} {et~al.}(2014){Judge}, {Kleint}, {Donea}, {Sainz Dalda}, \& {Fletcher}}]{Judge+others2014}
{Judge}, P.~G., {Kleint}, L., {Donea}, A., {Sainz Dalda}, A., \& {Fletcher}, L. 2014, ApJ, 796, 85

\bibitem[{{Kleint} {et~al.}(2020){Kleint}, {Berkefeld}, {Esteves}, {Sonner}, {Volkmer}, {Gerber}, {Kr{\"a}mer}, {Grassin}, \& {Berdyugina}}]{Kleint+others2020}
{Kleint}, L., {Berkefeld}, T., {Esteves}, M., {et~al.} 2020, A\&A, 641, A27, \dodoi{10.1051/0004-6361/202038208}

\bibitem[{{Klimchuk}(2006)}]{Klimchuk2006}
{Klimchuk}, J.~A. 2006, Solar Phys., 234, 41

\bibitem[{{Klimchuk} {et~al.}(2008){Klimchuk}, {Patsourakos}, \& {Cargill}}]{EBTEL}
{Klimchuk}, J.~A., {Patsourakos}, S., \& {Cargill}, P.~J. 2008, ApJ, 682, 1351, \dodoi{10.1086/589426}

\bibitem[{{Krieger} {et~al.}(1971){Krieger}, {Vaiana}, \& {van Speybroeck}}]{Krieger+others1971}
{Krieger}, A.~S., {Vaiana}, G.~S., \& {van Speybroeck}, L.~P. 1971, in Solar Magnetic Fields, ed. R.~{Howard}, Vol.~43 (Proc. IAU), 397--412

\bibitem[{{Kuin} \& {Martens}(1982)}]{Kuin+Martens1982}
{Kuin}, N.~P.~M., \& {Martens}, P.~C.~H. 1982, A\&A, 108, L1

\bibitem[{{Leenaarts} {et~al.}(2009){Leenaarts}, {Carlsson}, {Hansteen}, \& {Rouppe van der Voort}}]{Leenaarts+others2009}
{Leenaarts}, J., {Carlsson}, M., {Hansteen}, V., \& {Rouppe van der Voort}, L. 2009, ApJL, 694, L128

\bibitem[{{Leenaarts} {et~al.}(2016){Leenaarts}, {Golding}, {Carlsson}, {Libbrecht}, \& {Joshi}}]{Leenaarts+others2016}
{Leenaarts}, J., {Golding}, T., {Carlsson}, M., {Libbrecht}, T., \& {Joshi}, J. 2016, A\&A, 594, A104

\bibitem[{{Lemen} {et~al.}(2011){Lemen}, {Title}, {de Pontieu}, {Schrijver}, {Tarbell}, {et~al.}}]{IRIS}
{Lemen}, J., {Title}, A., {de Pontieu}, B., {et~al.} 2011, in AAS/Solar Physics Division Abstracts \#42, Vol. 289, 1512

\bibitem[{{Lemen} {et~al.}(2012){Lemen}, {Title}, {Akin}, {Boerner}, {Chou}, {Drake}, {Duncan}, {Edwards}, {Friedlaender}, {Heyman}, {Hurlburt}, {Katz}, {Kushner}, {Levay}, {Lindgren}, {Mathur}, {Mcfeaters}, {Mitchell}, {Rehse}, {Schrijver}, {Springer}, {Stern}, {Tarbell}, {Wuelser}, {Wolfson}, {Yanari}, {Bookbinder}, {Cheimets}, {Caldwell}, {Deluca}, {Gates}, {Golub}, {Park}, {Podgorski}, {Bush}, {Scherrer}, {Gummin}, {Smith}, {Auker}, {Jerram}, {Pool}, {Soufli}, {Windt}, {Beardsley}, {Clapp}, {Lang}, \& {Waltham}}]{Lemen+others2012}
{Lemen}, J.~R., {Title}, A.~M., {Akin}, D.~J., {et~al.} 2012, Sol. Phys., 275, 17, \dodoi{10.1007/s11207-011-9776-8}

\bibitem[{Linsky \& Avrett(1970)}]{Linsky+Avrett1970}
Linsky, J.~L., \& Avrett, E.~H. 1970, PASP, 82, 169

\bibitem[{{Litwin} \& {Rosner}(1993)}]{Litwin+Rosner1993}
{Litwin}, C., \& {Rosner}, R. 1993, ApJ, 412, 375

\bibitem[{{L{\'o}pez Fuentes} \& {Klimchuk}(2022)}]{Lopez+Klimchuk2022}
{L{\'o}pez Fuentes}, M., \& {Klimchuk}, J.~A. 2022, ApJ, 939, 17, \dodoi{10.3847/1538-4357/ac90c8}

\bibitem[{{Malanushenko} {et~al.}(2022){Malanushenko}, {Cheung}, {DeForest}, {Klimchuk}, \& {Rempel}}]{Malanushenko+others2022}
{Malanushenko}, A., {Cheung}, M.~C.~M., {DeForest}, C.~E., {Klimchuk}, J.~A., \& {Rempel}, M. 2022, ApJ, 927, 1

\bibitem[{{Mariska} {et~al.}(1982){Mariska}, {Doschek}, {Boris}, {Oran}, \& {Young}}]{Mariska+others1982}
{Mariska}, J.~T., {Doschek}, G.~A., {Boris}, J.~P., {Oran}, E.~S., \& {Young}, T.~R., J. 1982, ApJ, 255, 783, \dodoi{10.1086/159877}

\bibitem[{{Marsch} {et~al.}(2005){Marsch}, {Marsden}, {Harrison}, {Wimmer-Schweingruber}, \& {Fleck}}]{Orbiter}
{Marsch}, E., {Marsden}, R., {Harrison}, R., {Wimmer-Schweingruber}, R., \& {Fleck}, B. 2005, Advances in Space Research, 36, 1360

\bibitem[{{Marsden} {et~al.}(2013){Marsden}, {M{\"u}ller}, \& {StCyr}}]{Orbiter2}
{Marsden}, R.~G., {M{\"u}ller}, D., \& {StCyr}, O.~C. 2013, in American Institute of Physics Conference Series, Vol. 1539, Solar Wind 13, ed. G.~P. e.~a. {Zank}, 448--453

\bibitem[{{Martens}(2010)}]{Martens2010}
{Martens}, P.~C.~H. 2010, ApJ, 714, 1290, \dodoi{10.1088/0004-637X/714/2/1290}

\bibitem[{{McClymont} \& {Craig}(1985{\natexlab{a}})}]{McClymont+Craig1985a}
{McClymont}, A.~N., \& {Craig}, I.~J.~D. 1985{\natexlab{a}}, ApJ, 289, 820, \dodoi{10.1086/162946}

\bibitem[{{McClymont} \& {Craig}(1985{\natexlab{b}})}]{McClymont+Craig1985b}
---. 1985{\natexlab{b}}, ApJ, 289, 834, \dodoi{10.1086/162947}

\bibitem[{{M{\"u}ller} {et~al.}(2003{\natexlab{a}}){M{\"u}ller}, {Hansteen}, \& {Peter}}]{Muller+Hansteen+Peter2003}
{M{\"u}ller}, D.~A.~N., {Hansteen}, V.~H., \& {Peter}, H. 2003{\natexlab{a}}, A\&A, 411, 605

\bibitem[{{M{\"u}ller} {et~al.}(2003{\natexlab{b}}){M{\"u}ller}, {Biskamp}, \& {Grappin}}]{Mueller+others2003}
{M{\"u}ller}, W.-C., {Biskamp}, D., \& {Grappin}, R. 2003{\natexlab{b}}, \pre, 67, 066302

\bibitem[{Parker(1988)}]{Parker1988}
Parker, E.~N. 1988, ApJ, 330, 474

\bibitem[{{Patsourakos} \& {Klimchuk}(2005)}]{Patsourakos+Klimchuk2005}
{Patsourakos}, S., \& {Klimchuk}, J.~A. 2005, ApJ, 628, 1023, \dodoi{10.1086/430662}

\bibitem[{{Pesnell} {et~al.}(2012){Pesnell}, {Thompson}, \& {Chamberlin}}]{Pesnell+others2012}
{Pesnell}, W.~D., {Thompson}, B.~J., \& {Chamberlin}, P.~C. 2012, Sol. Phys., 275, 3

\bibitem[{{Pevtsov} {et~al.}(2003){Pevtsov}, {Fisher}, {Acton}, {Longcope}, {Johns-Krull}, {et~al.}}]{Pevtsov+others2003}
{Pevtsov}, A.~A., {Fisher}, G.~H., {Acton}, L.~W., {et~al.} 2003, ApJ, 598, 1387

\bibitem[{{Poletto} {et~al.}(1975){Poletto}, {Vaiana}, {Zombeck}, {Krieger}, \& {Timothy}}]{Poletto+others1975}
{Poletto}, G., {Vaiana}, G.~S., {Zombeck}, M.~V., {Krieger}, A.~S., \& {Timothy}, A.~F. 1975, Sol. Phys., 44, 83, \dodoi{10.1007/BF00156848}

\bibitem[{{Priest} {et~al.}(2002){Priest}, {Heyvaerts}, \& {Title}}]{Priest+others2002}
{Priest}, E.~R., {Heyvaerts}, J.~F., \& {Title}, A.~M. 2002, ApJ, 576, 533

\bibitem[{{Rachmeler} {et~al.}(2022){Rachmeler}, {Bueno}, {McKenzie}, {Ishikawa}, {Auch{\`e}re}, {Kobayashi}, {Kano}, {Okamoto}, {Bethge}, {Song}, {Ballester}, {Belluzzi}, {Pino Alem{\'a}n}, {Ramos}, {Yoshida}, {Shimizu}, {Winebarger}, {Kobelski}, {Vigil}, {Pontieu}, {Narukage}, {Kubo}, {Sakao}, {Hara}, {Suematsu}, {{\v{S}}t{\v{e}}p{\'a}n}, {Carlsson}, \& {Leenaarts}}]{Rachmeler+2022}
{Rachmeler}, L.~A., {Bueno}, J.~T., {McKenzie}, D.~E., {et~al.} 2022, ApJ, 936, 67, \dodoi{10.3847/1538-4357/ac83b8}

\bibitem[{{Rast} {et~al.}(2021){Rast}, {Bello Gonz{\'a}lez}, {Bellot Rubio}, {Cao}, {Cauzzi}, {Deluca}, {de Pontieu}, {Fletcher}, {Gibson}, {Judge}, {Katsukawa}, {Kazachenko}, {Khomenko}, {Landi}, {Mart{\'\i}nez Pillet}, {Petrie}, {Qiu}, {Rachmeler}, {Rempel}, {Schmidt}, {Scullion}, {Sun}, {Welsch}, {Andretta}, {Antolin}, {Ayres}, {Balasubramaniam}, {Ballai}, {Berger}, {Bradshaw}, {Campbell}, {Carlsson}, {Casini}, {Centeno}, {Cranmer}, {Criscuoli}, {Deforest}, {Deng}, {Erd{\'e}lyi}, {Fedun}, {Fischer}, {Gonz{\'a}lez Manrique}, {Hahn}, {Harra}, {Henriques}, {Hurlburt}, {Jaeggli}, {Jafarzadeh}, {Jain}, {Jefferies}, {Keys}, {Kowalski}, {Kuckein}, {Kuhn}, {Kuridze}, {Liu}, {Liu}, {Longcope}, {Mathioudakis}, {McAteer}, {McIntosh}, {McKenzie}, {Miralles}, {Morton}, {Muglach}, {Nelson}, {Panesar}, {Parenti}, {Parnell}, {Poduval}, {Reardon}, {Reep}, {Schad}, {Schmit}, {Sharma}, {Socas-Navarro}, {Srivastava}, {Sterling}, {Suematsu}, {Tarr}, {Tiwari}, {Tritschler}, {Verth}, {Vourlidas}, {Wang}, {Wang}, {NSO and DKIST
  Project}, {DKIST Instrument Scientists}, {DKIST Science Working Group}, \& {DKIST Critical Science Plan Community}}]{Rast+others2021}
{Rast}, M.~P., {Bello Gonz{\'a}lez}, N., {Bellot Rubio}, L., {et~al.} 2021, Sol. Phys., 296, 70, \dodoi{10.1007/s11207-021-01789-2}

\bibitem[{{Rimmele} {et~al.}(2020){Rimmele}, {Warner}, {Keil}, {Goode}, {Kn{\"o}lker}, {Kuhn}, {Rosner}, {McMullin}, {Casini}, {Lin}, {W{\"o}ger}, {von der L{\"u}he}, {Tritschler}, {Davey}, {de Wijn}, {Elmore}, {Fehlmann}, {Harrington}, {Jaeggli}, {Rast}, {Schad}, {Schmidt}, {Mathioudakis}, {Mickey}, {Anan}, {Beck}, {Marshall}, {Jeffers}, {Oschmann}, {Beard}, {Berst}, {Cowan}, {Craig}, {Cross}, {Cummings}, {Donnelly}, {de Vanssay}, {Eigenbrot}, {Ferayorni}, {Foster}, {Galapon}, {Gedrites}, {Gonzales}, {Goodrich}, {Gregory}, {Guzman}, {Guzzo}, {Hegwer}, {Hubbard}, {Hubbard}, {Johansson}, {Johnson}, {Liang}, {Liang}, {McQuillen}, {Mayer}, {Newman}, {Onodera}, {Phelps}, {Puentes}, {Richards}, {Rimmele}, {Sekulic}, {Shimko}, {Simison}, {Smith}, {Starman}, {Sueoka}, {Summers}, {Szabo}, {Szabo}, {Wampler}, {Williams}, \& {White}}]{Rimmele+others2020}
{Rimmele}, T.~R., {Warner}, M., {Keil}, S.~L., {et~al.} 2020, Sol. Phys., 295, 172, \dodoi{10.1007/s11207-020-01736-7}

\bibitem[{{Rochus} {et~al.}(2020){Rochus}, {Auch{\`e}re}, {Berghmans}, {Harra}, {Schmutz}, {Sch{\"u}hle}, {Addison}, {Appourchaux}, {Aznar Cuadrado}, {Baker}, {Barbay}, {Bates}, {BenMoussa}, {Bergmann}, {Beurthe}, {Borgo}, {Bonte}, {Bouzit}, {Bradley}, {B{\"u}chel}, {Buchlin}, {B{\"u}chner}, {Cab{\'e}}, {Cadiergues}, {Chaigneau}, {Chares}, {Choque Cortez}, {Coker}, {Condamin}, {Coumar}, {Curdt}, {Cutler}, {Davies}, {Davison}, {Defise}, {Del Zanna}, {Delmotte}, {Delouille}, {Dolla}, {Dumesnil}, {D{\"u}rig}, {Enge}, {Fran{\c{c}}ois}, {Fourmond}, {Gillis}, {Giordanengo}, {Gissot}, {Green}, {Guerreiro}, {Guilbaud}, {Gyo}, {Haberreiter}, {Hafiz}, {Hailey}, {Halain}, {Hansotte}, {Hecquet}, {Heerlein}, {Hellin}, {Hemsley}, {Hermans}, {Hervier}, {Hochedez}, {Houbrechts}, {Ihsan}, {Jacques}, {J{\'e}r{\^o}me}, {Jones}, {Kahle}, {Kennedy}, {Klaproth}, {Kolleck}, {Koller}, {Kotsialos}, {Kraaikamp}, {Langer}, {Lawrenson}, {Le Clech'}, {Lenaerts}, {Liebecq}, {Linder}, {Long}, {Mampaey}, {Markiewicz-Innes}, {Marquet},
  {Marsch}, {Matthews}, {Mazy}, {Mazzoli}, {Meining}, {Meltchakov}, {Mercier}, {Meyer}, {Monecke}, {Monfort}, {Morinaud}, {Moron}, {Mountney}, {M{\"u}ller}, {Nicula}, {Parenti}, {Peter}, {Pfiffner}, {Philippon}, {Phillips}, {Plesseria}, {Pylyser}, {Rabecki}, {Ravet-Krill}, {Rebellato}, {Renotte}, {Rodriguez}, {Roose}, {Rosin}, {Rossi}, {Roth}, {Rouesnel}, {Roulliay}, {Rousseau}, {Ruane}, {Scanlan}, {Schlatter}, {Seaton}, {Silliman}, {Smit}, {Smith}, {Solanki}, {Spescha}, {Spencer}, {Stegen}, {Stockman}, {Szwec}, {Tamiatto}, {Tandy}, {Teriaca}, {Theobald}, {Tychon}, {van Driel-Gesztelyi}, {Verbeeck}, {Vial}, {Werner}, {West}, {Westwood}, {Wiegelmann}, {Willis}, {Winter}, {Zerr}, {Zhang}, \& {Zhukov}}]{EUI}
{Rochus}, P., {Auch{\`e}re}, F., {Berghmans}, D., {et~al.} 2020, A\&A, 642, A8, \dodoi{10.1051/0004-6361/201936663}

\bibitem[{Rosner(1990)}]{Rosner1990}
Rosner, R. 1990, in {Physics of Magnetic Flux Ropes}, ed. C.~T. {Russell}, E.~R. {Priest}, \& L.~C. {Lee}, Vol.~58 (Washington DC American Geophysical Union Geophysical Monograph Series), 189, \dodoi{10.1029/GM058}

\bibitem[{Rosner {et~al.}(1978)Rosner, Tucker, \& Vaiana}]{Rosner+Tucker+Vaiana1978}
Rosner, R., Tucker, W.~H., \& Vaiana, G.~S. 1978, ApJ, 220, 643

\bibitem[{{Schmidt} {et~al.}(2012){Schmidt}, {von der L{\"u}he}, {Volkmer}, {Denker}, {Solanki}, {Balthasar}, {Bello Gonz{\'a}lez}, {Berkefeld}, {Collados}, {Fischer}, {Halbgewachs}, {Heidecke}, {Hofmann}, {Kneer}, {Lagg}, {Nicklas}, {Popow}, {Puschmann}, {Schmidt}, {Sigwarth}, {Sobotka}, {Soltau}, {Staude}, {Strassmeier}, \& {Waldmann}}]{GREGOR2012}
{Schmidt}, W., {von der L{\"u}he}, O., {Volkmer}, R., {et~al.} 2012, Astronomische Nachrichten, 333, 796, \dodoi{10.1002/asna.201211725}

\bibitem[{{Schrijver} \& {Harvey}(1989)}]{Schrijver+harvey1989}
{Schrijver}, C.~J., \& {Harvey}, K.~L. 1989, ApJ, 343, 481

\bibitem[{{Schrijver} \& {Zwaan}(2000)}]{Schrijver+Zwaan2000}
{Schrijver}, C.~J., \& {Zwaan}, C. 2000, {Solar and Stellar Magnetic Activity} ({New York}: Cambridge University Press)

\bibitem[{{Title} \& {Schrijver}(1998)}]{Title+Schrijver1998}
{Title}, A.~M., \& {Schrijver}, C.~J. 1998, in Astronomical Society of the Pacific Conference Series, Vol. 154, Cool Stars, Stellar Systems, and the Sun, ed. R.~A. {Donahue} \& J.~A. {Bookbinder}, 345

\bibitem[{{Toriumi} \& {Airapetian}(2022)}]{Toriumi+Airapetian2022}
{Toriumi}, S., \& {Airapetian}, V.~S. 2022, ApJ, 927, 179, \dodoi{10.3847/1538-4357/ac5179}

\bibitem[{{Uzdensky}(2007)}]{Uzdensky2007}
{Uzdensky}, D.~A. 2007, ApJ, 671, 2139, \dodoi{10.1086/522915}

\bibitem[{{van den Oord} \& {Barstow}(1988)}]{Vandenoord+Barstow1988}
{van den Oord}, G.~H.~J., \& {Barstow}, M.~A. 1988, A\&A, 207, 89

\bibitem[{{van Speybroeck} {et~al.}(1970){van Speybroeck}, {Krieger}, \& {Vaiana}}]{Vanspeybrock+Krieger+Vaiana1970}
{van Speybroeck}, L.~P., {Krieger}, A.~S., \& {Vaiana}, G.~S. 1970, Nat., 227, 818

\bibitem[{Vernazza {et~al.}(1981)Vernazza, Avrett, \& Loeser}]{Vernazza+Avrett+Loeser1981}
Vernazza, J., Avrett, E., \& Loeser, R. 1981, ApJS, 45, 635

\bibitem[{{Wang}(2016)}]{Wang2016}
{Wang}, Y.~M. 2016, ApJL, 820, L13, \dodoi{10.3847/2041-8205/820/1/L13}

\bibitem[{{Wentzel}(1978)}]{Wentzel1978}
{Wentzel}, D.~G. 1978, SP, 58, 307

\end{thebibliography}

\end{document}